\documentclass[preprint,flushrt,tighten]{aastex}
\usepackage{bbm}

\slugcomment{}

\shorttitle{Are Coronae Heated by Flares? III.}
\shortauthors{Arzner and G\"udel}

\begin{document}

\title{Are Coronae of Magnetically Active Stars Heated by Flares? \\
    III. Analytical Distribution of Superimposed Flares}

\author{Kaspar Arzner and Manuel G\"udel}
\affil{Paul Scherrer Institut, CH-5232 Villigen PSI, Switzerland}
\email{arzner@astro.phys.ethz.ch, guedel@astro.ethz.ch }

\begin{abstract}
 
We study the hypothesis that observed X-ray/extreme ultraviolet emission from
coronae of magnetically active stars is entirely (or to a large part) due to 
the superposition of flares, using an analytic approach to determine the 
amplitude distribution of flares in light curves. The flare-heating
hypothesis is motivated by time series that show continuous variability
suggesting the presence of a large number of superimposed flares with similar 
rise and decay time scales. We rigorously relate the amplitude distribution of 
stellar flares to the observed histograms of binned 
counts and photon waiting times, under the assumption that the flares occur at random and have similar shapes.
Our main results are: (1) The characteristic function (Fourier transform of the probability density) of the
expected counts in time bins $\Delta t$ is $\phi_F(s,\Delta t)$ = $\exp \{ - T^{-1} \hspace{-1mm} \int_{-\infty}^\infty
\hspace{-2mm} dt \, [1 - \phi_a (s \, \Xi (t,\Delta))] \}$, where $T$ is the mean flaring interval, $\phi_a(s)$ is the
characteristic function of the flare amplitudes, and $\Xi (t, \Delta t)$ is
the flare shape convolved with the observational time bin. 
(2) The probability of finding $n$ counts in time bins $\Delta t$ is
$P_c(n)$ = $(2\pi)^{-1} \hspace{-1mm} \int_0^{2\pi} \hspace{-1mm}ds \, e^{-ins} \phi_F(s,\Delta t)$.
(3) The probability density of photon waiting times $x$ is $P_\delta(x)$ = 
$\partial_x^2 \phi_F(i,x) / \langle r \rangle$ with $\langle r \rangle$ = $\partial_x \phi_F(i,x)|_{x=0}$
the mean count rate. An additive independent background is readily included.
Applying these results to EUVE/DS observations of the flaring star AD Leo,
we find that the flare amplitude distribution can be represented by a truncated power law
with a power law index of 2.3 $\pm$ 0.1. Our analytical results 
agree with existing Monte Carlo results of \cite{kashyap02} and \cite{guedel03}.
The method is applicable to a wide range of further stochastically bursting astrophysical
sources such as cataclysmic variables, Gamma Ray Burst substructures, X-ray binaries,
and spatially resolved observations of solar flares.
   
\end{abstract}


\section{INTRODUCTION}

Since the late 1930s the solar corona has been known to be much hotter than the photosphere, 
although the latter delineates the surface through which energy enters the corona. 
The corona is thus not in a state of simple heat flow equilibrium, and its high
temperature must be supplied with some form of mechanical, electrical or radiative work which 
is dissipated at coronal heights. Several heating mechanisms have been proposed, such as
Alfv\'en \citep{Hollweg, HeyvaertsPriest83, MarshTu} or ion cyclotron \citep{AxfordMcKenzie}
waves, or electric currents \citep{HeyvaertsPriest84} and magnetic reconnection \citep{Parker72} 
converting magnetic shear into jets, and, ultimately, into heat. Some of these 
processes operate in a continuous mode, whereas others invoke transients. Comprehensive reviews may be found,
e.g., in \citet*{narain90}, \citet*{ulmschneider91}, \cite{benz95}, \cite*{narain96},
and \cite{PriestForbes}.

With increasing sensitivity of the observations it was recognized
that the `quiet' solar corona contains in fact `microflares' and `nanoflares' down to the actual resolution limit. 
Clearly, these events are the signature of non-equilibrium processes, and therefore represent promising candidates for 
the primary heating mechanism of the solar corona \citep{kruckerbenz98, benzkrucker99}. 

Stars at high magnetic activity levels show properties that are reminiscent
of continuous flaring processes, such as persistent coronal temperatures of
5--30 MK,  high coronal electron densities, or continuous non-thermal radio
emission. See \citet{guedel03} for a summary
of observational results. Foremostly, X-ray and extreme ultraviolet (EUV) time
series of such stars reveal a high level of flux variability on time scales
similar to flares \citep*{butler86, ambrust87, collura88, kashyap99,
audard99, audard00, osten99}. Limited detector sensitivity could
give the wrong impression that the emission is due to steadily emitting coronal
sources, while new, highly sensitive observations in at least some
cases show little indication of a base level emission that could be defined
as being genuinely `quiescent' \citep{audard03,guedel02}. A coronal heating mechanism that
should successfully explain X-ray and EUV emission from magnetically active stars
inevitably has to address the question of continuous variability.
 
In the context of microflare heating one is interested in the frequency distribution of flare 
amplitudes. In particular, it is essential
to have reliable estimators of the small but frequent flares, since these presumably
carry a major part of the total energy. However, these small flares substantially 
overlap, so that their amplitudes are hard to determine from the noisy observations. 
One way out is to use flare identification algorithms that also include flare superposition \citep{audard00}.
Another way out is to resort to a forward model, which constrains the flare amplitude distribution
by model assumptions. Within a Monte Carlo framework, this approach has been followed
by \cite{kashyap02} and in \cite{guedel03}.

In the present article we explore {\it analytical} forward methods. We describe a statistical model of
superimposed flares which provides an exact relation between the (univariate) probability
density of the flare amplitudes and the (univariate) probability
densities of the observed binned count rates and photon waiting times. 
As the method involves the histograms of binned count rates and photon
waiting times only, it is insensitive to permutation of the time bins and to the presence of long
data gaps. The price for this robustness is the loss of temporal information, which is replaced by 
the assumption that all flares have similar shape and occur at random.
See \cite{Isliker96} and \cite{kraevbenz} for a complementary approach focusing 
on the temporal autocorrelation of the light curves (and thus on the first two moments of the 
univariate count rate distribution).

In what follows we basically work with the observables `counts' and `count rates', 
with emphasis on the effect of flare superposition and counting noise.
The conversion into luminosity and flare energy is briefly discussed in Section 
\ref{Discussion_Sect}.

\section{\label{ModelSect} MODEL}

\begin{figure}[h!]
\plotone{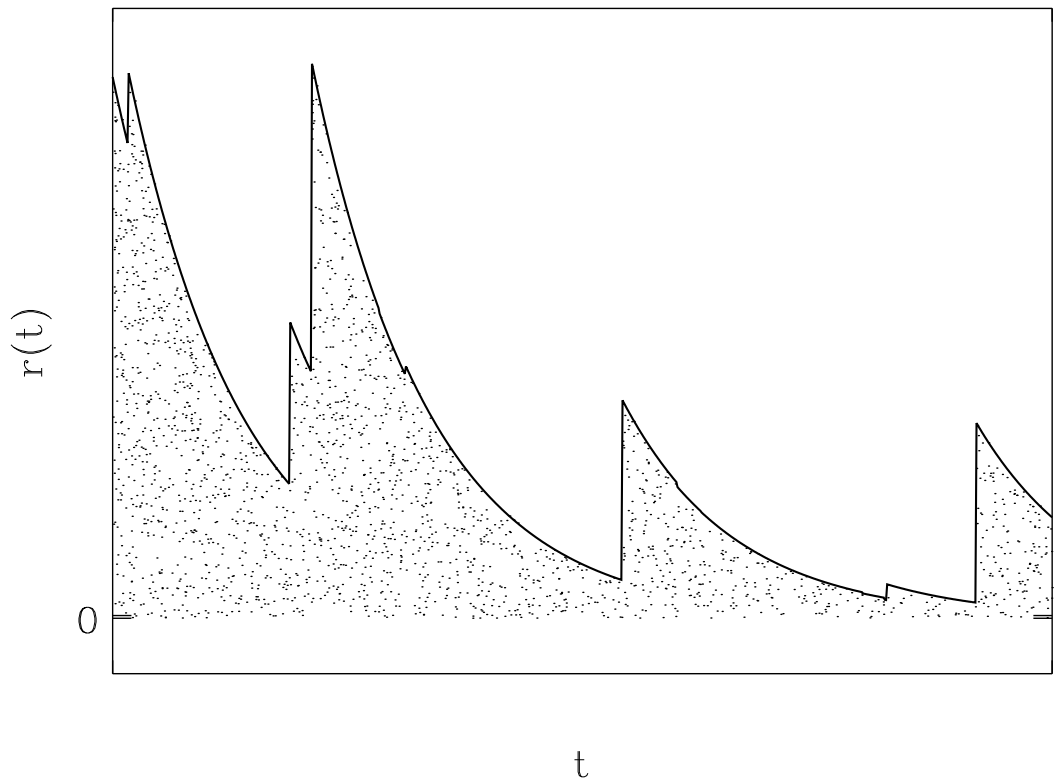}
\vspace{-10mm}
\caption{Construction of the time-dependent Poisson process with intensity $r(t)$ from
the standard 2D Poisson process: imagine the whole 2D plane uniformly covered with points 
of density $\rho$ = $\langle r \rangle$ $\times$ 1s. By definition,
the probability of finding {\it no} point in an arbitrary domain of area $A$ is $e^{-\rho A}$.
Points below the graph $r(t)$ (boldface) are again a Poisson process, and so is their projection onto the $t$ axis. 
The resulting $t$-coordinates are identified with the photon arrival times belonging to $r(t)$.}
\label{PoissonProcess_fig}
\end{figure}

While the energy release and relaxation processes 
are not known in detail, there is observational support for the similarity of flares
on a global scale -- at least for `simple' flares. Stellar flares of the type investigated in this 
paper can usually well be represented by a sharp rise and an exponential decay. Examples supporting 
this view are found in \cite{dejager89}, \cite{guedel02}, \cite{guedel04}, and \cite{audard03}.
The observed durations of flares, as investigated in the solar case, are almost independent of the flare
amplitude or radiated energy even if 3-5 observed orders of magnitude are included.
(\citealt{pearce88, shimizu95, feldman97, aschwanden00},
see also summary in \citealt{guedel03}). The decay times of flares studied below
were found to scatter around a constant value in \cite{guedel03}, independent of
peak flux. Since in stellar observations 3 orders of magnitude of flare energy are usually
sufficient to explain the observed emission, we shall assume that a constant decay time is
a good approximation for our case as well. For completeness, the case of a weak dependence of the decay
time scale $\tau$ on radiated total energy $E$, $\tau \propto E^{0.25}$ as
suggested by one stellar investigation, was studied in both \citet{kashyap02}
and \cite{guedel03}, with the result that the main conclusions on the importance
of small flares were reinforced. Observations also suggest that individual flares occur
at random and independently of each other (for the solar case, see \citealt{wheatland00,wheatland03}). 
This independence might suggest that flares trigger an exhaustive `reset' where mechanical stress (and thus
memory) is completely destroyed. We shall thus adopt the following model for the 
instantaneous count rate in a fixed energy band:

\begin{equation}
r(t) = \sum_k a_k \, \xi(t - t_k) + b(t) \;\;\;\;\;\;\; \mbox{[ct s$^{-1}$]}
\label{model}
\end{equation}

where $\xi (t)$ represents the (dimensionless) flare shape, $a_k$ [ct s$^{-1}$] are the flare amplitudes,
and $b(t)$ [ct s$^{-1}$] is a background intensity. We shall agree that
$0 \le \xi (t) \le \xi (0)=1$, and define $\tau = \int_{-\infty}^\infty \hspace{-1mm} \xi (t)dt$ to be the 
flare duration. All flare amplitudes $a_k$ are positive and drawn from the same probability density $P_a(x)$.
(From here on, $x$ denotes a generic real positive variable.)
The flare times $t_k$ occur at random with mean flaring interval $T$ [s], and it is assumed that all $a_k$, 
$t_k$, and the background are statistically independent of each other. The observed counts are supposed to be
a Poisson process with time-varying intensity $r(t)$. Such a non-uniform Poisson process is not unambiguously
defined, and we use here the construction shown in Fig. \ref{PoissonProcess_fig}: the photon arrival times are 
distributed like the $t$-components of uniformly distributed points $(t,y)$ in the $ty$-plane
satisfying $y < r(t)$. Note that the background is included in $r(t)$.
When binned in time bins of duration $\Delta t$, the counts are Poisson distributed with
parameter (=expected number of counts in $\Delta t$)

\begin{equation}
R(t,\Delta t) = \int_t^{t+\Delta t} \hspace{-6mm} dt' \, r(t') \, .
\label{R}
\end{equation}

Alternatively, we may work with the photon event list. In this case we consider the time intervals
between consecutive photon arrivals (`photon waiting times').

Dimensional analysis shows that the flare-related part of the signal (\ref{model}) is governed by the single 
dimensionless parameter $\tau / T$. The conversion into discrete time bins 
involves a second dimensionless parameter, $\Delta t / \tau$ (or, alternatively, $\Delta t / T$).
If the peaks are well-separated ($\tau / T \ll 1$) and 
counting noise is low, one can directly estimate the flare amplitudes from the observed peaks in the
light curve. If flares overlap ($\tau / T > 1$) or counting noise is relevant, this is no longer
possible. We resort then to the model (\ref{model}) and ask for its predictions on the observed 
probability densities of binned counts $P_c(n)$ and photon waiting times $P_\delta(x)$,
which are available with high accuracy under stationary conditions. The goal is thus
to establish the relation between $P_a(x)$ on the one hand, and $P_c(n)$
and $P_\delta(x)$ on the other hand.

\section{\label{GenSolution}SOLUTION}

The linearity of Equation (\ref{model}), together with the statistical independence of its ingredients, 
suggests that a rigorous mapping from $P_a(x)$ to $P_c(n)$ and $P_\delta(x)$ should exist. This 
is in fact the case, and the individual steps of this mapping are discussed in Sections 3.1 - 3.4. 
See Table 1 for an overview on our notation.

\begin{table}[h!]
\caption{Notation$^a)$}
\begin{tabular}{|clc|}\hline
$\xi (t)$		& flare shape &			\\
$\tau$		& flare duration [s], $\tau=\int_{-\infty}^\infty \hspace{-2.5mm} dt \, \xi (t)$ & \\
$T$		& mean flaring interval [$s$] 	& \\	
$\Delta t$	& observational time bin [s]	& \\
$\Xi (t,\Delta t)$	& = $\int_t^{t+\Delta} \hspace{-1mm} dt' \, \xi(t')$ &	\\\hline\hline
probability density & quantity $x,n$ & characteristic function\\\hline
$P_a(x)$ 	& flare amplitudes [ct~$s^{-1}$]	& $\phi_a(s)$ 	\\
$P_f(x)$	& flare intensity [ct~$s^{-1}$] & $\phi_f(s)$	\\
$P_F(x)$	& flare parameter [ct] & $\phi_F(s,\Delta t)$ \\
$P_b(x)$	& background intensity [ct~$s^{-1}$]	& $\phi_b(s)$	\\
$P_B(x)$	& background parameter [ct] 	& $\phi_B(s,\Delta t)$	\\
$P_r(x)$	& total intensity [ct~$s^{-1}$]	& $\phi_r(s)$	\\
$P_R(x)$	& total parameter [ct]	& $\phi_R(s,\Delta t)$	\\
$P_c(n)$	& binned counts	& $\phi_c(s,\Delta t)$	\\
$P_\delta(x)$	& photon waiting time [s]	&		\\\hline
\end{tabular}

{$^a)$\small The (Poisson-) `intensity' is the instantaneous photon count rate; the 
(Poisson-) `parameter' is the expected number of photons in time bins $\Delta t$. `Total'
means the sum of background and flares.}
\end{table}

\subsection{\label{flare_contrib_Sect}Flare Contributions}

We first ask for the probability that the flare signal $f(t) = \sum_k a_k \xi (t-t_k)$ assumes values
between $x$ and $x + dx$. By statistical homogeneity, all times are equivalent.
Consider thus the time $t=0$, and let it be centered in an auxiliary
interval $I$ of length $|I|$. Neglecting flares outside $I$
as well as the background, the intensity at $t$ = 0 is $f(0) = \sum_{t_k \in I} a_k \, \xi (-t_k)$. 
Since $f(0)$ is a sum of independent random variates, it is conveniently \citep{Levy} represented in
a way which diagonalizes convolution, such as by its Laplace- or Fourier transform.
We use here the latter, called the characteristic function \citep{Lukacs} and defined by
\begin{equation}
\phi_f(s) \doteq \langle e^{isf(0)} \rangle = \int_0^\infty \hspace{-3mm} dx \, P_f(x) \, e^{isx}
\label{phi_define}
\end{equation}
where $P_f(x) \, dx$ is the probability that $f(0)$ is between $x$ and $x+dx$. The probability density is obtained from 
the characteristic function by Fourier back transform, $P_f(x) = (2\pi)^{-1} \int \phi_f(s) e^{-isx} \, ds$.
We continue now with the calculation of the characteristic function $\phi_f(s)$. The essential point to notice is 
that $\phi_f(s)$ can be partitioned according to the number of flares which occur in the interval $I$. One can thus 
write $\phi_f(s) = \sum_{l=0}^\infty P_l \, \langle e^{isf_l} \rangle$, where 
$P_l = e^{-|I|/T} (|I|/T)^l/l!$ is the probability of finding exactly $l$ 
flares in $I$, and $f_l = \sum_{k=1}^l a_k \xi(-t_k)$ 
is the corresponding photon intensity from these $l$ superimposed flares in $I$. 
It is understood that $f_0=0$, because the case $l$ = 0 flare does not contribute to 
the count rate. Since the times $t_k$ 
are independent and uniformly distributed in $I$, the characteristic function of $f_l$ is
\begin{eqnarray}
\langle e^{isf_l} \rangle & = & \int_0^\infty \hspace{-2mm} d a_1 \, P_a(a_1) \;  ... \int_0^\infty \hspace{-2mm} da_l \, P_a(a_l) \; 
		\frac{1}{|I|} \int_I \hspace{-1mm} dt_1 \; ... \; \frac{1}{|I|} \int_I \hspace{-1mm} dt_l 
			\, \exp \Big\{ i s \sum_{k=1}^l a_k \xi(-t_k) \Big\} \label{phi_q1} \\
		& = & \frac{1}{|I|^l} \int_I \hspace{-1mm} dt_1 \; ... \, \int_I \hspace{-1mm} dt_l		
		 \, \phi_a \Big(s \, \xi(-t_1) \Big) \cdot ... \cdot \phi_a \Big(s \, \xi(-t_l) \Big) \label{phi_q2} \\
		& = & \frac{1}{|I|^l} \bigg( \int_I \hspace{-1mm} d t \, \phi_a \Big( s \,
			 \xi (t) \Big) \bigg)^l \, . \label{phi_q3}
\end{eqnarray}
%
%
The step from equation (\ref{phi_q1}) to equation (\ref{phi_q2}) also invokes the 
fact that all $a_k$ are independently $P_a$-distributed,
and the definition of the characteristic function of the flare amplitudes, $\phi_a(s)$
= $\langle e^{isa} \rangle$. The step from equation (\ref{phi_q2})
to equation (\ref{phi_q3}) makes use of the symmetry of $I$ about $t$ = 0.
Performing the sum over all possible numbers of flares and using equation (\ref{phi_q3}), one finds that
\begin{equation}
\phi_f(s) = \langle e^{isf(0)} \rangle = \sum_{l=0}^\infty \frac{e^{-|I|/T} (|I|/T)^l}{l \, !} \langle e^{isf_l} \rangle
	= \exp \Big\{ - T^{-1} \hspace{-1mm} \int_I dt \, \big[1 - \phi_a \big( s \, \xi (t) \big) \big] \Big\} \label{phi_fT} \, .
\end{equation}
We now want to liberate ourselves from the artificial time interval $I$, by extending it to infinity and thereby collecting all
flare contributions. To this end we have to investigate whether the integral in equation (\ref{phi_fT}) remains finite as $|I| \to \infty$.
This turns out to be the case under rather weak conditions, one of which is as follows (see Appendix \ref{limit_Appendix}): if for some $\alpha > 0$ the
flare shape $\xi(t)$ decays faster than $|t|^{-1/\alpha}$ as $|t| \to \infty$, and if the probability density of the flare amplitudes 
decays no slower than $P_a(x) \sim x^{-1-\alpha}$ as $x \to \infty$, then equation (\ref{phi_fT}) remains well-behaved
as $|I| \to \infty$. Note that the above condition does not require the expectation value
of the flare amplitudes to be finite; this freedom will allow us to consider
every power law index $\nu > 1$ if the flare amplitudes have a power law distribution truncated at some lower cutoff.
($\nu > 1$ must be required to ensure that $P_a(x)$ can be normalized.)
Letting thus $I$ go to infinity, equation (\ref{phi_fT}) becomes 
\begin{equation}
\phi_f(s) = \exp \Big\{ - T^{-1} \hspace{-1mm} \int_{-\infty}^{\infty} \hspace{-3mm} dt \,
	\big[1 - \phi_a \big( s \, \xi (t) \big) \big] \Big\} \, .
\label{phi_f}
\end{equation}
The argumentation leading to equation (\ref{phi_f}) can be repeated, with minor modifications, for the
Poisson parameter (expected counts) of time bins of duration $\Delta t$. Neglecting the background, the
Poisson parameter of time bin $[0, \Delta t]$ is $F(0,\Delta t) \doteq \int_0^{\Delta t} \hspace{-1mm} dt \, \sum_k a_k \xi (t-t_k)
= \sum_k a_k \int_{-t_k}^{-t_k+\Delta t} \hspace{-2mm} dt' \, \xi (t')$, 
so that we merely have to replace $\xi (t)$ by
\begin{equation}
\Xi (t,\Delta t) \doteq \int_t^{t+\Delta t} \hspace{-4mm} dt' \, \xi (t')
\label{Xi(t)}
\end{equation}
in order to obtain the characteristic function of $F(0,\Delta t)$,
\begin{equation}
\phi_F(s,\Delta t) = \langle e^{isF(0,\Delta t)} \rangle = \exp \Big\{ - T^{-1} \hspace{-1mm} \int_{-\infty}^{\infty} \hspace{-3mm} dt \, \
	\big[1 - \phi_a \big( s \, \Xi (t,\Delta t) \big) \big] \Big\} \, .
\label{phi_F}
\end{equation}
Equations (\ref{phi_f},\ref{phi_F}) are our central result; they establish the connection between 
the probability density of flare amplitudes $P_a$ via $\phi_a$
and the distributions of photon intensities and expected counts, $P_f$, $P_F$ via $\phi_f$, $\phi_F$. 
Equation (\ref{phi_F}) holds under the same conditions as equation (\ref{phi_f}).
The function $\Xi (t,\Delta t)$ is a blurred version of the flare shape 
with resolution degraded to the observational time bin. It is easily seen that probabilistic normalization 
is preserved ($\phi_f(0)$=1 and $\phi_F(0,\Delta t)$=1 if $\phi_a(0)$=1). It can also be shown that non-negative
definiteness \citep{Lukacs} of $\phi_a(s)$ is inherited under the transformations (\ref{phi_f}) and (\ref{phi_F}), 
so that the \cite{Bochner} theorem ensures that the Fourier inverses of $\phi_f(s)$ and $\phi_F(s,\Delta t)$ represent, 
in fact, probability densities. The transformation $P_a(x) \to P_F(x)$ redistributes amplitudes
$x$ both towards higher (flare overlap) and lower (flare tails) values. It also broadens the probability density
function by finite time binning, unless $\Delta t \ll \tau$, in which case $\Xi (t,\Delta t) \to \Delta t \, \xi (t)$
and $\phi_F(s,\Delta t) \to \phi_f(s \, \Delta t )$. In the limit $\Delta t \gg T$ and
if $\langle a^2 \rangle < \infty$, normal 
behavior is approached, with $F(0,\Delta t)/\Delta t$ concentrated about $\langle a \rangle \tau / T$ 
with relative spread of order $(\Delta t / T)^{-1/2}$. This expected result may be proven
by formal expansion of equation (\ref{phi_F}) about $1/\Delta t = 0$.

\subsection{Background}

So far the background has been neglected. When included, the total (flare plus background) photon Poisson 
intensity at time $t$ = 0 is $r(0) = f(0)+b(0)$, and the total Poisson parameter of the time bin $[0,\Delta t]$ 
is $R(0,\Delta t) = F(0,\Delta t) + B(0,\Delta t)$ with $B(0,\Delta t) = \int_0^{\Delta t} \hspace{-1mm} b(t) \, dt$. Since the background is independent 
of the flares, the characteristic functions of the total photon Poisson intensity and -parameter are
\begin{eqnarray}
\phi_r(s) & = & \phi_b(s) \, \phi_f(s) \label{phi_r} \\
\phi_R(s,\Delta t) & = & \phi_B(s,\Delta t) \, \phi_F(s,\Delta t) \label{phi_R}
\end{eqnarray}
with $\phi_b(s)$ and $\phi_B(s,\Delta t)$ the characteristic functions of $b(0)$ and $B(0,\Delta t)$, respectively. If the
background is constant then $\phi_B(s,\Delta t) = \phi_r(s\Delta t) = e^{isb\Delta t}$.

\subsection{Binned Counts\label{binned_Sect}}

The counts in the time bin [$0,\Delta t$] are Poisson distributed with parameter $R(0,\Delta t)$,
where $R(0,\Delta t)$ itself is $P_R$-distributed. The binned counts have thus the characteristic function
\begin{equation}
\phi_c(s,\Delta t) = \langle e^{isn} \rangle = \sum_{n=0}^\infty \int dx \, P_R(x) \frac{e^{-x} x^n}{n!} e^{isn}
	= \int dx \, P_R(x) \, e^{ix(i-ie^{is})} = \phi_R(i-ie^{is} \hspace{-1mm},\Delta t) \, ,
\label{phi_c}
\end{equation}
which is $2\pi$-periodic because the binned counts are integers.
The probability density of the binned counts is obtained from Fourier back transform of 
$\phi_c(s,\Delta t)$; when expressed in terms of the original probability density $P_a(x)$, 
the probability of finding $n$ counts in bins of duration $\Delta t$ is
\begin{equation}
P_c(n) = \frac{1}{2\pi} \hspace{-1mm} \int_0^{2 \pi} \hspace{-4mm} \phi_B(s,\Delta t) \, \exp \Big\{-ins - \frac{1}{T} \hspace{-1mm}
	\int_{-\infty}^\infty \hspace{-4mm} dt \int_0^\infty \hspace{-3mm} dx \, P_a(x) 
	\Big(1 - e^{-x(1-e^{is})\Xi (t,\Delta t)} \Big) \Big\} \, .
\label{P_c}
\end{equation}
If a Gaussian approximation was used, with the binned counts modeled by $c = R(0,\Delta t) + \sqrt{R(0,\Delta t)} \, X$ 
with $X$ standard normal, then $c$ could assume any real number and the characteristic function is
$\phi_c(s,\Delta t) \simeq \phi_R(s+is^2/2,\Delta t)$. This approximation becomes exact at
$s \to 0$, corresponding to large count rates.

\subsection{\label{WaitingTimeSect} Photon Waiting Times}

Instead of time-binning the observed counts one may directly work with the photon event list, and
consider the histogram of waiting times $\delta t$ between consecutive photon arrivals. This has the advantage that no
ad-hoc choice of time bins is needed, and that the waiting time bins can be adapted for optimum 
(similar) probability content. In addition, long data gaps ($\gg \delta t$) and detector saturation ($\delta t \ge \tau_{sat}$) 
are easily segregated, and the construction of the photon waiting time histogram amounts to a
direct determination of the moment generating function of $r(t)$, which is (almost) the
quantity provided by theory. The photon waiting times represent therefore a natural interface
between theory and observation.

To see why this is so, consider the `waiting' event that there is (i) one photon between $t_1$ and $t_1+dt$,
(ii) no photon between $t_1+dt$ and $t_2$, and (iii) one photon between $t_2$ and $t_2 + dt$ (the waiting
time is thus $t_2-t_1 > 0$). By the binomial theorem, the probabilities of (i) and (iii) are $r(t_1)\,dt + O(dt^2)$ 
and $r(t_2)\,dt + O(dt^2)$, while from Figure \ref{PoissonProcess_fig} we see that the probability of (ii) is $\exp \{ - \int_{t_1}^{t_2} \hspace{-1mm} r(t) \, dt \}$.
Since (i), (ii) and (iii) are independent of each other,
the waiting event occurs with probability density $r(t_1) \exp \{-\int_{t_1}^{t_2} \hspace{-1mm} r(t) \, dt \} 
\,r(t_2) = - \partial_{t_1} \partial_{t_2} \, \exp \{-\int_{t_1}^{t_2} \hspace{-1mm} r(t) \, dt \}$. Denoting the average
over realizations of $r(t)$ by angular brackets, and using the translation invariance $\langle \exp \{-\int_{t_1}^{t_2} \hspace{-1mm} r(t) 
\,dt \} \rangle = \langle \exp -\int_0^{t_2-t_1}\hspace{-1mm}r(t)\,dt \rangle$, one finds that the probability density
of photon waiting times $x = t_2 - t_1$ is
\begin{equation}
P_\delta(x) = N \, \partial_x^2 \, \langle e^{-\int_0^x \hspace{-1mm} r(t) \, dt} \rangle 
= \frac{1}{\langle r \rangle} \frac{\partial^2 \phi_R(i,x)}{\partial x^2} \, ,
\label{P_d}
\end{equation}
where the normalization constant $N$ is determined by the constraint $\int_0^\infty \hspace{-1mm} P_\delta(x) \, dx$ = 1,
implying\footnote{Because of $\int_0^\infty \hspace{-1mm} P_\delta(x) \, dx = N \, \partial_x \langle \exp - \int_0^x \hspace{-1mm} r(t) 
\, dt \rangle \big|_0^\infty = - N \, \lim_{x \to 0} \partial_x \langle \exp- \int_0^x \hspace{-1mm} r(t) \, dt \rangle = N 
\langle r \rangle$.} that $N = 1/\langle r \rangle$. While equation (\ref{P_d}) is exact, it is generally difficult to
calculate $\phi_R(i,x)$ for arbitrary $x$. This is, however, not needed because the photon waiting time is 
much shorter than the flare duration in any practically relevant case. One may thus resort to the approximation 
$\phi_R(i,x) \simeq \phi_r(ix)$ and
\begin{equation}
P_\delta(x) \simeq - \frac{\phi_r''(ix)}{\langle r \rangle} \;\;\;\; \mbox{if} \;\;\;\; x \ll \tau \;\;\; 
\mbox{and} \;\;\; \langle a \rangle \tau \gg 1 \, . \label{P_d0} 
\end{equation}
The cumulative waiting time distribution is, accordingly, 
\begin{equation}
\int_0^x \hspace{-2mm} P_\delta(y) \, dy \simeq 1 + i \frac{\phi_r'(ix)}{\langle r \rangle} \;\;\;\; \mbox{if} \;\;\;\; x \ll \tau \;\;\; 
\mbox{and} \;\;\; \langle a \rangle \tau \gg 1 \,\, .
\label{P_d1}
\end{equation}
If the background is time-varying on time scale $\tau_b$,
the conditions $x \ll \tau_b$ and $\langle b \rangle \tau_b \gg 1$ must be added to the right hand side of 
equations (\ref{P_d0}, \ref{P_d1}). The range of validity of equation (\ref{P_d0}) was explored analytically 
(Appendix \ref{explicit_Appendix}) and by Monte Carlo simulations (Section \ref{Simulations}). 
It was found to hold even if the assumptions on the right hand side are not strictly valid.

The result (\ref{P_d0}) can also be derived in the following, perhaps more intuitive, way:
let $P_r(y) \, dy$ denote the probability that $r(t)$ is between $y$ and $y+dy$ 
in a random time interval of fixed duration. Then, $y P_r(y)/\int y' P_r(y') dy'$ is the 
probability that a given photon has been emitted at Poisson parameter $y$. (This probability
is proportional to the total area of strips in the $(t,y)$ plane with $y \le r(t) \le y+dy$
-- see Fig. \ref{PoissonProcess_fig}). Assume now that
$r(t)$ varies only weakly during the local waiting time $1/r(t)$. Thus 
$y P_r(y)/\int y' P_r(y') dy'$ is also the probability that a waiting time `belongs'
to intensity $y$. Given the intensity $y$, the photon waiting time has probability density 
$y \, e^{-y \, \delta t}$, so that the (marginal) probability density of photon waiting times $x$ is
$P_\delta(x) = \int dy \, P_r(y) \, y^2 e^{-yx} / \int dy \, P_r(y) \, y = 
- \phi_r''(i \, x) / \langle r \rangle$, which is equation (\ref{P_d0}).

\subsection{Special Results}

For special flare shapes and amplitude distributions, equations (\ref{phi_f}) - (\ref{P_d1})
can be evaluated rigorously. This section summarizes results which are relevant for application to
observational data (Section \ref{obs_Sect} below).

\subsubsection{Flare Shape}

Exponential decay is a generic feature of many relaxation processes, and is a natural model of
flares. Observations of late-type main sequence stars often reveal a rather
abrupt onset followed by an approximately exponential decay (for stellar examples, see 
\citealt{guedel02, guedel03, audard03}), with decay time almost independent of flare
size (e.g., \citealt{shimizu95}; see also Section II). We shall thus use an exponential flare shape with constant
decay time $\tau$ as a proxy for exact calculations. If the flare shape is 
$\xi (t) = \theta(t) \, e^{-t/\tau}$ with $\theta(t)$ the Heaviside step function, then
\begin{equation}
\phi_f(s) = \exp \Big\{ - \frac{\tau}{T} \int_0^s \hspace{-1mm} d y \, \frac{1-\phi_a(y)}{y} \Big\} 
\label{phi_f1}
\end{equation}
(see Appendix \ref{explicit_Appendix}). The abrupt and artificial onset of the one-sided exponential flare shape may 
be softened by an exponential increase, thus creating a two-sided exponential shape: 
$\xi (t) = \frac{1}{2}\theta(-t)e^{t/\tau^+}+\frac{1}{2}\theta(t)e^{-t/\tau^-}$. 
Since flare contributions from $t>0$ and $t<0$ to $\xi (0)$ are statistically independent of each other, 
the characteristic function of their combined flare photon intensity is
\begin{equation}
\phi_f(s) = \phi^-_f(s) \, \phi^+_f(s) \, ,
\label{phi_f_pm}
\end{equation} 
where $\phi_f^\pm(s)$ is given by equation (\ref{phi_f1}) with $\tau$ replaced by $\tau^\pm$. Equations 
(\ref{phi_f1}) and (\ref{phi_f_pm}) show that two-sided exponential flares have the
same characteristic function $\phi_f$ as one-sided exponential flares with effective decay time
$\tau^-+\tau^+$. A similar result holds for $\phi_F$ if $\Delta t \ll \tau$.
From the observational point of view, this implies that one- and two-sided exponential 
flares cannot be distinguished based on their short-term photon waiting time (or binned count)
distribution. We therefore assume in the sequel that the flare shape is one-sided
exponential, with the understanding that our results on $P_a(x)$ are robust against
the generalization of a one-sided to two-sided exponential shapes.

\subsubsection{Flare Amplitude Distribution}

For several forms of $P_a(x)$ the characteristic function $\phi_f(s)$
can be explicitly obtained. Solvable cases include exponentials, power laws, and L\'evy
distributions. Details are given in Appendix \ref{explicit_Appendix}; here we cite 
the results for power laws, which are a traditional and observationally well confirmed model of 
the distributions of solar \citep*{datlowe74, lin84, kruckerbenz98, benzkrucker99, aschwanden00} 
and stellar \citep{collura88, audard99, audard00} flares. If $P_a(x)$ is a power law 
with exponent $\nu$, truncated at a lower cutoff $x=a_0$,
\begin{equation}
P_a(x) = (\nu-1) \, a_0^{\nu-1} \, \theta(x-a_0) \, x^{-\nu} \, ,
\label{P_a_pl}
\end{equation}
and if the flare shape is exponential, then $\phi_f(s)$ can be exactly expressed in 
terms of special functions (Appendix \ref{explicit_Appendix}). 
For practical purposes, these may be approximated by elementary functions. Of particular
interest is the cumulative photon waiting time distribution (eq. \ref{P_d1}) which
predicts the probability content of arbitrary photon waiting time bins:
\begin{eqnarray}
\int_0^x \hspace{-1mm} P_\delta(y) \, dy & \simeq & 1 + \bigg\{ -b + \lambda\tau a_0(\nu-1)
	\Big( C (a_0x)^{\nu-2} - \frac{1}{\nu-2} 
	+ \frac{a_0x}{2\nu-6} - \frac{(a_0x)^2}{9\nu-36} \Big) \bigg\} \times \nonumber \\
      & & \times  \frac{ \exp \bigg\{ - bx + \lambda\tau a_0x \Big( C(a_0x)^{\nu-2}
      - \frac{\nu-1}{\nu-2} + \frac{(\nu-1)a_0x}{4\nu-12} \Big) \bigg\} } 
      { b + \frac{\lambda\tau a_0(\nu-1)}{\nu-2} }
\label{Pd_approx}
\end{eqnarray}
with $C = \pi [\Gamma(\nu) \sin (\pi \nu)]^{-1}$. The relative accuracy of the approximation
(\ref{Pd_approx}) was found to be better than 0.001 if $a_0 \,x \le 0.25$ and $\nu$ $>$ 2.01. 
For $\nu$ $>$ 2 the expectation value of the flare amplitudes exists and is given by
$\langle a \rangle = a_0 (\nu-1)/(\nu-2)$. In the limit $\nu$ $\to$ 2 the right hand
side of equation (\ref{Pd_approx}) converges to the unit step function $\theta(x)$; this is also
true if the exact equations (\ref{phi_f_pl1}, \ref{phi_f_pl2}) are used.
The limit $\langle a \rangle$ $\to$ $\infty$ therefore predicts, theoretically,
arbitrarily small photon waiting times.

\section{VALIDATION OF RESULTS}

Before application to real data, the results of Section \ref{GenSolution} have been verified 
by several independent methods. The introduction of the characteristic function in Section 
\ref{flare_contrib_Sect} may appear somewhat arbitrary, and the question arises whether a more 
direct approach exists. The answer is yes, but the calculation is tedious 
and difficult to make rigorous. We shall therefore only sketch it in Appendix \ref{moment_Appendix}. 
Here, we discuss validation based on an entirely independent analytical argument
(Section \ref{Peak_Sect}) and on numerical simulations (Section \ref{Simulations}).

\subsection{\label{Peak_Sect}The Distribution of Consecutive Peaks}

For one-sided exponential flares -- and only for these -- 
the signal $\sum_k a_k \xi (t-t_k)$ is a continuous-time Markov process whose stationary
measure can be obtained from a renewal argument. This provides an independent verification 
of Equation (\ref{phi_f1}), and has some interest on its own as it allows to relax the Poisson assumption 
on the flare times (see Sect. \ref{Discussion_Sect} below).The argument is as follows. 
Discarding the background, the amplitudes at consecutive peaks are related by
\begin{equation}
r_n = a_n + r_{n-1} e^{-\eta_n/\tau}
\label{rek1}
\end{equation}
with $r_n = r(t_n+0)$ and $\eta_n = t_n-t_{n-1}$ the flare waiting times. Now, in the stationary case, the 
probability densities of $r_n$ and $r_{n-1}$ must agree. Therefore their characteristic function $\phi_{r_n}$ must satisfy
\begin{equation}
\phi_{r_n}(s) = \langle e^{isr_n} \rangle = \langle e^{ia_ns} \rangle \langle e^{i s r_{n-1} e^{-\eta_n/\tau} } \rangle 
= \phi_a(s) \int_0^\infty \hspace{-2mm} P_\eta (x) \, \phi_{r_n} (s e^{-x/\tau}) \, d x \, ,
\label{phi_eq}
\end{equation}
where $P_\eta (x)$ is the flare waiting time distribution, and we have used the definition of the characteristic 
function and the fact that $a_n$, $r_n$ and $\eta_n$ are statistically independent of
each other. For uniformly distributed flares, the flare waiting times are exponentially
distributed, $P_\eta (x) = T^{-1} e^{-x/T}$. Upon setting $z$ = $ \, e^{-\eta/\tau}$, equation (\ref{phi_eq}) 
simplifies thus to 
\begin{equation}
\phi_{r_n}(s)= \frac{\tau}{T} \phi_a(s) s^{-\tau / T} \int_0^s z^{-1+\tau/T} \phi_{r_n}(z)\, dz \, .
\label{int_eq}
\end{equation}
Equation (\ref{int_eq}) can be solved for $\phi_{r_n}(s)$ by taking 
derivatives with respect to $s$, yielding
\begin{equation}
\phi_{r_n}'(s) = \frac{\phi_a'(s)}{\phi_a(s)} \phi_{r_n}(s) - \frac{\tau}{T} \frac{\phi_{r_n}(s)}{s} 
+ \frac{\tau}{T} \phi_a(s) \frac{\phi_{r_n}(s)}{s} \, .
\label{diff_eq}
\end{equation}
Together with the initial (normalization) condition $\phi_{r_n}(0) = 1$, equation (\ref{diff_eq}) has 
the unique solution
\begin{equation}
\phi_{r_n}(s) = \phi_a(s) \, \exp \Big\{ - \frac{\tau}{T} \int_0^s \frac{1-\phi_a(y)}{y} \, dy  \Big\} \, .
\label{phi_r_n}
\end{equation}
%
%
The exponential factor on the right hand side of equation (\ref{phi_r_n}) acts as a low-pass filter, which
broadens $P_{r_n}(x)$ with respect to $P_a(x)$ by an amount proportional to $\tau / T$. As it must be, 
isolated flares reproduce the true flare amplitudes, $\lim_{\tau / T \to 0} \phi_{r_n}(s)=\phi_a(s)$.
There is an apparent similarity between equations (\ref{phi_f1}) and (\ref{phi_r_n}), and the exact
relation is as follows. By construction, the peaks $r_n$ are the sum of two independent random variates: the value of the
continuous signal just before the jump, $r(t_n-0)$, plus the flare amplitude $a_n$. From equation (\ref{phi_r_n})
we see that $r(t_n-0)$ has the characteristic function $\exp \{ - \tau T^{-1} \hspace{-1mm} \int_0^s (1-\phi_a(x))x^{-1} \, dx \}$.
Since $r(t_n-0)$ occurs with equal probability than any $r(t)$, the above expression must agree with the result of 
equation (\ref{phi_f1}). This is in fact the case.

\subsection{\label{Simulations}Simulations}

Equations (\ref{phi_f}, \ref{phi_F}, \ref{P_c}, \ref{P_d}) have also been verified by Monte Carlo simulations. 
In these simulations, a realization of $r(t)$ is generated, and the binned Poisson parameters 
$R(t,\Delta t)$ = $\int_t^{t+\Delta t} \hspace{-2mm} r(t') \, dt'$ are numerically calculated. The binned counts
are then simulated using a conventional Poisson number generator, and their histogram is created.
For the photon waiting times we proceed as indicated in Figure \ref{PoissonProcess_fig}: a large number 
$N = YT$ of points $(t,y)$ are uniformly distributed in the rectangle
$[0,T] \times [0,Y]$, with $T \sim 3 \times 10^3 \, \tau$ and $Y \sim \max \, r(t)$.
The $t$-coordinates of points with $y < r(t)$ constitute our photon list. Note that the background
is included in $r(t)$. The simulations are compared to numerical evaluations of eqns. (\ref{P_c},\ref{P_d},\ref{P_d0}) 
and to exact results (Appendix \ref{explicit_Appendix}).

\begin{figure}[h!]
\epsscale{0.85}
\plotone{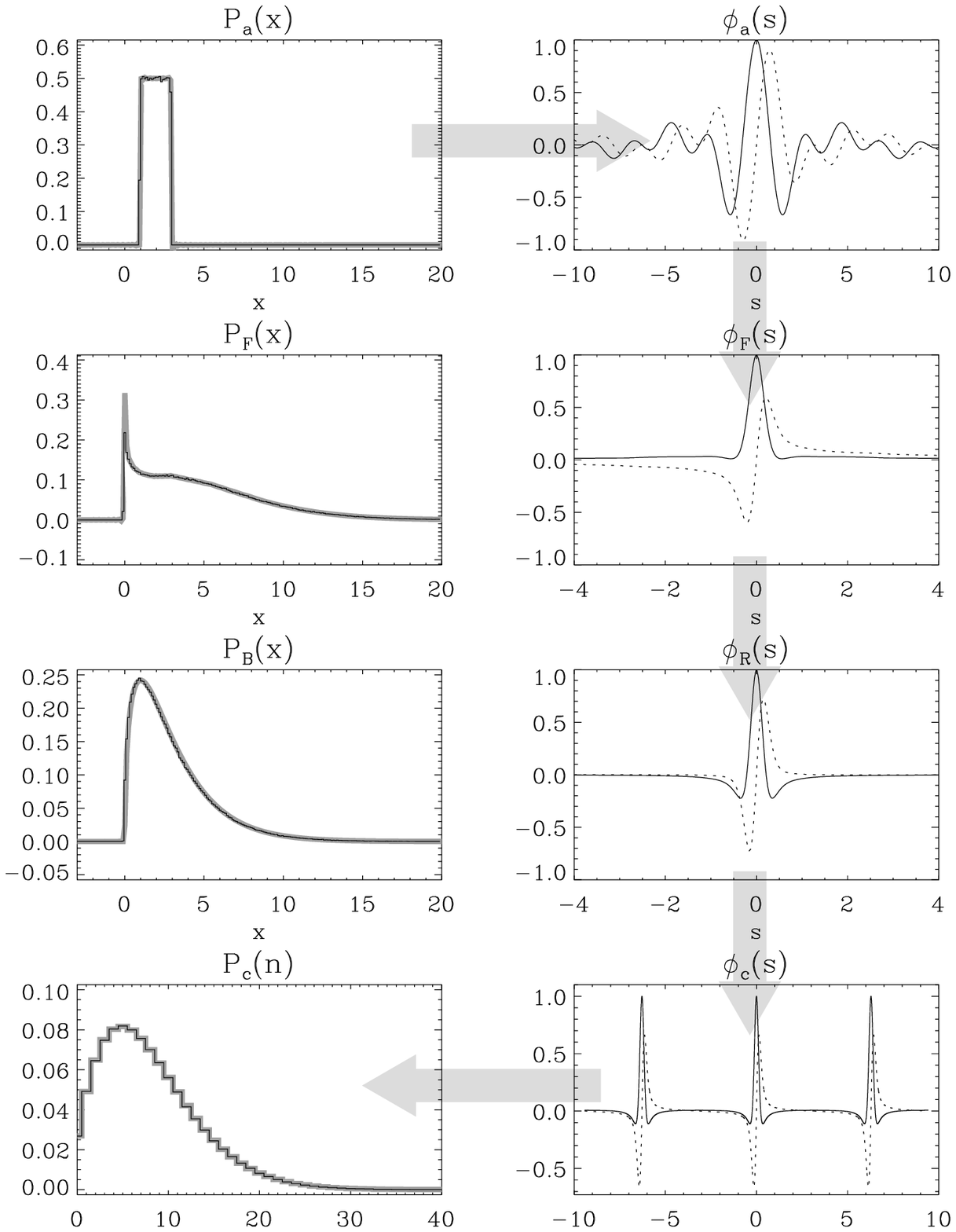}
\caption{The transformation from a (rectangular) flare 
amplitude probability density (top left) to the probability density of the binned counts (bottom left). Black (gray) line indicate 
simulation (theory). The flare shape is exponential ($\tau / T$=0.8) and the bin size is 
$\Delta t = 0.375\tau$. Line 2 (left) shows the probability density of the flare photon Poisson 
parameter; line 3 (left) shows the probability density of the binned background. The right 
column shows the real (solid) and imaginary (dashed) parts of the corresponding characteristic 
functions, except for line 3 (right) which refers to total (flare plus background) photon 
Poisson parameter. Arrows indicate the flow of the transformation $P_a(x) \to P_c(n)$.}
\label{binned_counts_fig}
\end{figure}

Figure \ref{binned_counts_fig} (left column) shows the theoretical (gray) and simulated (black) 
probability densities which occur in the transformation from the flare amplitudes into the binned counts.
The flare amplitudes are uniformly distributed between 1 and 3 ct~$s^{-1}$, the flare profile is a one-sided
exponential with decay time $\tau$ = 8 s, the mean flaring interval is $T$ = 10 s, and 
the simulation involves $2 \times 10^6$ time bins of duration $\Delta t$ = 3 s. 
These parameters represent marginally resolved, partially overlapping flares.
Accordingly, $P_F(x)$ (Fig. \ref{binned_counts_fig} line 2 left) is smoothed with respect to 
$P_a(x)$ (line 1 left) but has also a peak at $x$=0 which is due to low-count sequences between different flares.
The binned background $B(0,\Delta t)$ is modeled as chi square distributed with 3 degrees of freedom 
(Fig. \ref{binned_counts_fig} line 3 left); this ad-hoc choice yields a particularly simple characteristic
function and is for illustration purposes only. The conversion of the binned Poisson
parameter (Fig. \ref{binned_counts_fig} line 3 right) into discrete counts 
renders the characteristic function $2\pi$-periodic (Fig. \ref{binned_counts_fig} line 4 right). 
For benchmarking purposes, the theoretical probability densities are obtained from 
numerical Fourier inversion of the associated characteristic functions (Fig. \ref{binned_counts_fig} 
right column, solid line - real part, dashed line - imaginary part).

\begin{figure}[h!]
\epsscale{0.8}
\plotone{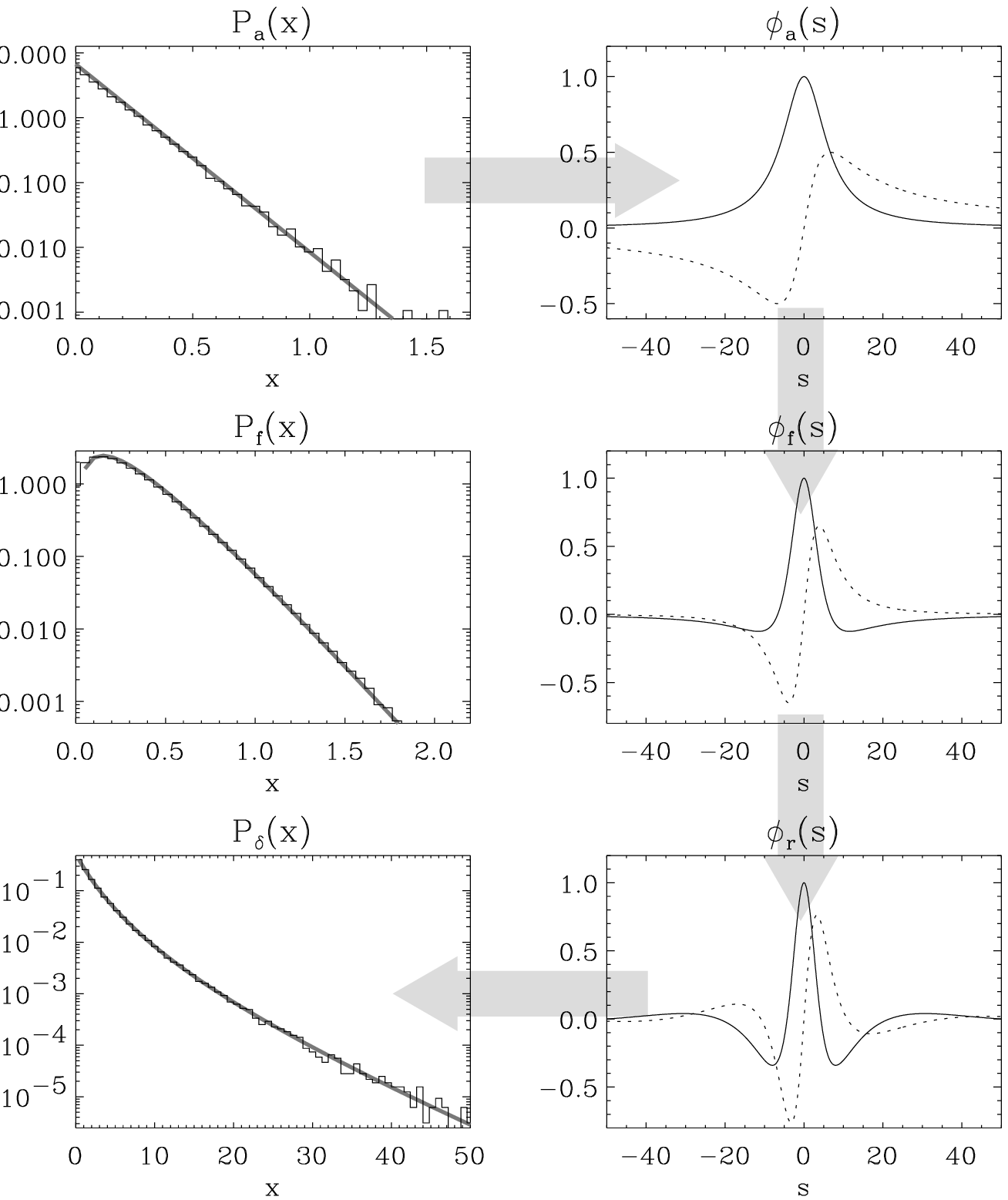}
\caption{Similar to Figure \protect\ref{binned_counts_fig}, but for the photon waiting times.
Left column: simulated (black) and theoretical (gray) probability densities of
flare amplitudes (top), flare photon intensity (middle) and photon waiting times (bottom). 
Right column: characteristic functions of flare amplitudes (top), flare  
photon intensity (middle) and flare-plus-background photon intensity (bottom). The flare amplitudes are exponentially 
distributed with mean $a$=0.15 ct~$s^{-1}$, the flare shape is one-sided exponential with $\tau$=40 s, the mean flaring interval 
is $T$=20 s, and there is a constant background of 0.1 ct~$s^{-1}$. The simulation involves 
3$\times$10$^5$ photons. Arrows indicate the flow of the transformation $P_a(x) \to P_\delta(x)$.}
\label{waitingtime_fig}
\end{figure}

The transformation from the flare amplitudes into the photon waiting times is illustrated in Figure
\ref{waitingtime_fig}. The flare amplitudes are exponentially distributed (mean amplitude $a$ = 0.15 ct~$s^{-1}$,
mean flaring interval $T$ = 20 s) and the flare shape is one-sided exponential ($\tau$ = 40 s). The
exponential flare amplitudes are chosen because they admit simple analytical
results for $P_\delta(x)$ (see Appendix \ref{explicit_Appendix}), which can be compared to the simulation.
A constant background $b$ = 0.1 ct~$s^{-1}$ is present, and the simulated photon list includes 
$N \sim 3 \times 10^5$ photons. The theoretical $P_f(x)$ (Fig. \ref{waitingtime_fig} middle 
left gray) is a chi square distribution with $\tau / T$ degrees of freedom (eq. \ref{phi_f_exp}), and 
$P_\delta(x)$ (Fig. \ref{waitingtime_fig} bottom left gray) is given by the exact
equation (\ref{P_d_exp_bg}). The characteristic functions of the
flare amplitudes, flare photon intensity, and total (flare plus background) photon intensity 
are shown in the right column of Figure \ref{waitingtime_fig}. Solid line denotes the real 
part, and dashed line denotes the imaginary part. A broad probability density
corresponds to narrow characteristic functions, and vice versa. The constant background amounts to
a modulation by $e^{isb}$.

\begin{figure}[h!]
\plotone{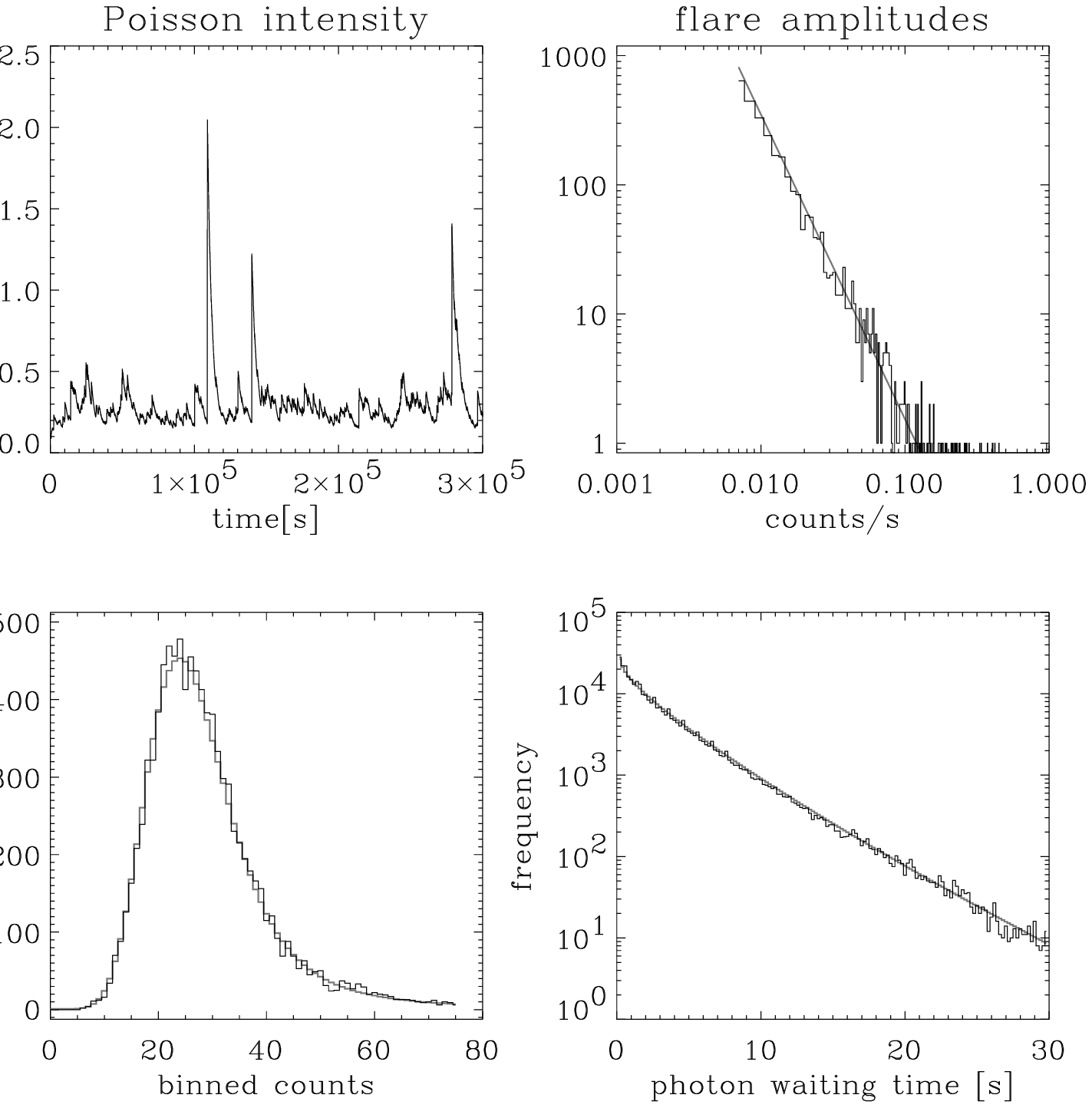}
\caption{Simulation: realization of $r(t)$ (top left) of flares with one-sided
exponential shape ($\tau$=3000 s) and power law distributed amplitudes (top right), together
with the histograms of binned counts and photon waiting times. Black (gray) line denotes
simulation (theory).}
\label{sim_pl_fig}
\end{figure}

Both binned counts and photon waiting times have also been simulated for power law flare
amplitudes, which are used as a fitting template for the EUVE/DS observations below.
Figure \ref{sim_pl_fig} shows the result of a simulation of one-sided exponential flares
($\tau$ = 3000 s, $T$ = 333 s) with lower cutoff $a_0$ = 0.007 ct~$s^{-1}$ and power
law index $\nu$ = 2.36. The simulation involves 3$\times 10^5$ photons in
10$^4$ time bins of duration $\Delta t$ = 100 s. Again, the black line represents simulation,
and the gray line represents theory, with $P_\delta(x)$ being the short-term form (eq. \ref{P_d0})
computed by differentiation of the exact expression (\ref{phi_f_pl1}) for $\phi_f(s)$.
A constant background $b$ = 0.09 ct~$s^{-1}$ is present, and an upper 
cutoff $a_1$ = 7 ct~$s^{-1}$ has been applied in the simulations and when
evaluating $P_c(n)$ from equation (\ref{P_c}).

\begin{figure}[h!]
\epsscale{0.95}
\plotone{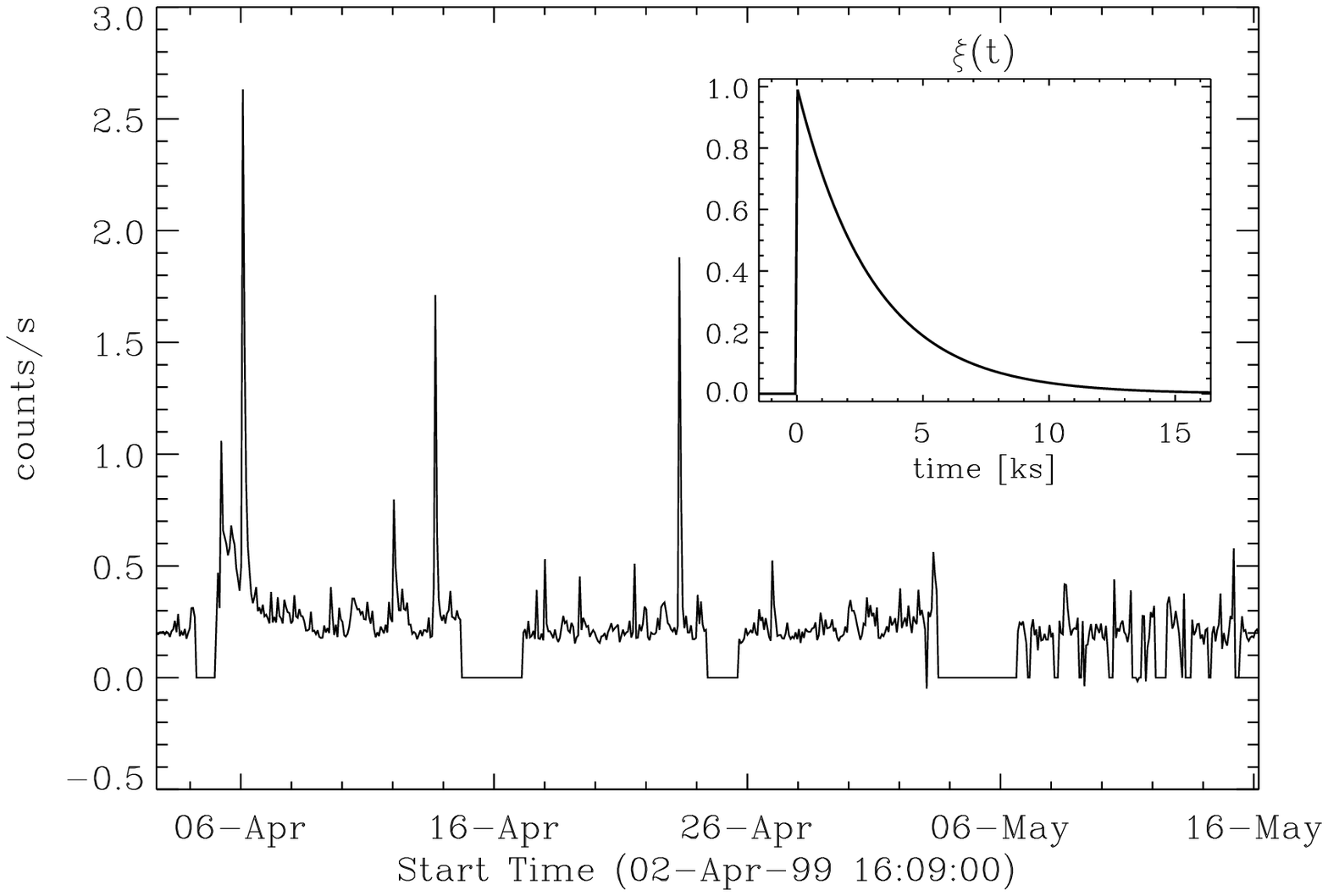}
\vspace{5mm}
\caption{EUVE/DS light curve from AD Leo obtained between April 2 and May 16 1999,
integrated over one spacecraft orbit (5663 s). The model flare shape is shown as inlet, adopting the
decay time of 3000 s of \protect\cite{kashyap02}.}
\label{EUVE_lc_fig}
\end{figure}

\section{\label{obs_Sect}OBSERVATIONS AND NUMERICAL RESULTS}

We have applied our theoretical results to a 3700 ks (elapsed time) observation of
the active star AD Leo acquired by the Extreme Ultraviolet Explorer (EUVE, e.g. \citealt{MalinaBowyer})
Deep Survey (DS) instrument between April 2 and May 15 1999. This time series has previously been analyzed 
by \cite{kashyap02} and in \cite{guedel03} by Monte-Carlo simulations 
of the model (eq. \ref{model}), so that we can compare our findings with existing results.
The Deep Survey instrument records photons between 52 \AA \, and 246 \AA \, 
with maximum sensitivity at 91 \AA. Figure \ref{EUVE_lc_fig} shows an overview of the 
light curve, time-integrated over spacecraft orbits (5663 s each).
During one spacecraft orbit, AD Leo was in the EUVE field of view for typically 1500 s, 
followed by a $\sim$ 4200 s interval without observations. Furthermore, there are observation interrupts 
where the light curve of Figure \ref{EUVE_lc_fig} drops to zero.
The period from 05/06 16:42 to 05/16 05:11 carries large contamination from energetic particles and
is excluded from our analysis. 

\begin{figure}[h!]
\plotone{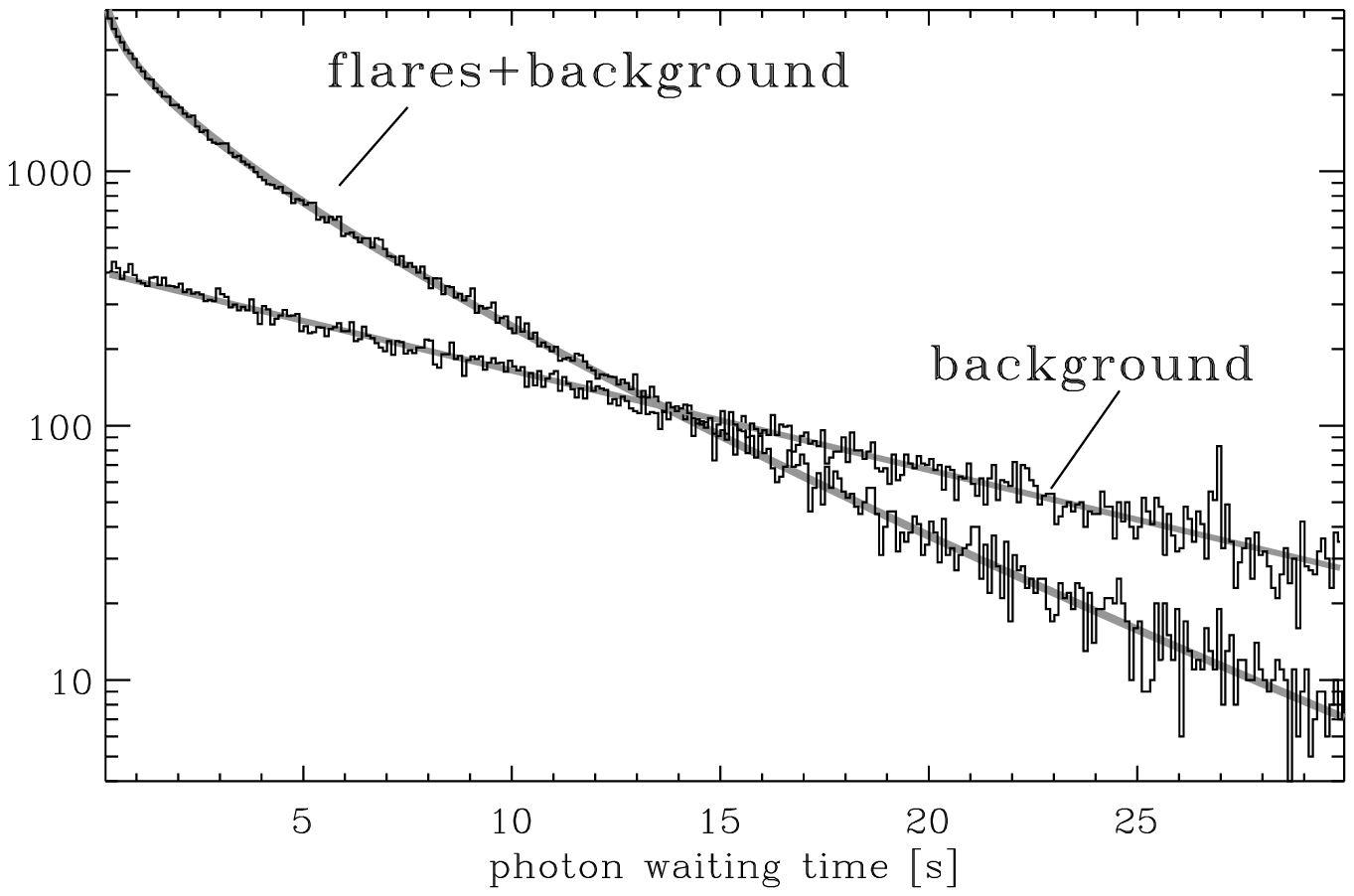}
\vspace{6mm}
\caption{Observed (black) and theoretical (gray) histograms of the photon waiting times
of the EUVE/DS AD Leo observation (Fig. \protect\ref{EUVE_lc_fig}). The waiting time bin
size is 0.1 s, the background is modeled as constant, and the flare amplitudes are modeled 
by a truncated power law.}
\label{EUVE_Pd_fig}
\end{figure}

Figure \ref{EUVE_Pd_fig} shows the histograms of photon waiting times. The `flares+background'
are collected across a spatial window centered at AD Leo, while the `background'
is determined from a larger disjoint area. 
The resolution of our photon arrival times is 0.1 s. This is not the full EUVE/DS time resolution, 
but has been considered sufficient with regard to the actual count rates. Only photon waiting times 
from 0.25 s and 30 s are taken into account. The lower limit eliminates poor flare statistics
and telemetry saturation at rare and extremely bright flares, and the upper limit eliminates low count rates
and artifacts from (much longer) unobserved intervals.
Our selection comprises 118,637 photon waiting times, and the photon waiting time bin size is 0.1 s.
The background waiting times (Fig. \ref{EUVE_Pd_fig} `background' black line) are found to closely follow 
an exponential distribution (Fig. \ref{EUVE_Pd_fig} `background' gray line); accordingly, we model the background 
as constant. When scaled to the `flares+background' collecting area, the best-fit background intensity 
is (cf. \citealt{kashyap02})
\begin{equation}
b = 0.03 \, \pm 0.002 \; \mbox{ct~$s^{-1}$} \, .
\label{b}
\end{equation}
Once the background is known, we determine the flare amplitude distribution $P_a(x)$ under the
assumption that $P_a(x)$ is a power law with exponent $\nu$ and lower cutoff $a_0$ (eq. \ref{P_a_pl}),
and that the flare shape is one-sided exponential (Fig. \ref{EUVE_lc_fig} inlet). These 
assumptions correspond -- up to a little relevant upper cutoff -- to those made by 
\cite{kashyap02} and in \cite{guedel03}. We also assert that the
short-term forms (eqns. \ref{P_d0}, \ref{P_d1}) of the photon waiting time distribution apply.
Our model parameters are thus ($a_0, \, \nu, \, \tau / T$). They are determined by minimizing
\begin{equation}
\chi^2 = \sum_n \frac{\big(H_n-P_n\big)^2}{P_n} \, ,
\label{chi2}
\end{equation}
where $H_n$ ($P_n$) is the observed (predicted) number of waiting times in the $n$-th waiting time bin.
The prediction is given by $P_n = N \int_{x_n}^{x_{n+1}} \hspace{-1mm} P_\delta(x) \, dx \big/ \int_{0.25s}^{30s} 
\hspace{-1mm} P_\delta(x) \, dx$, and we use the exact expression (\ref{phi_f_pl1}) when computing
the cumulative photon waiting time distribution (eq. \ref{P_d1}).
The sum in equation (\ref{chi2}) runs over $N_o$ = $(30-0.25)/0.1$ = 298 waiting time bins, and there are 
$N_{\rm DOF}$ = $N_o$ $-$ 3 = 295 degrees of freedom. The minimum reduced chi square is 0.989, 
and the best-fit parameters are
\begin{equation}
\nu = 2.29 \pm 0.07 \;\;\;\;\;\; a_0 = 0.0049 \pm 0.0015 \; \mbox{ct~$s^{-1}$} \;\;\;\;\;\; \tau / T = 11 \pm 4 \, .
\label{results_d}
\end{equation}
The resulting theoretical waiting time distribution is shown in Figure \ref{EUVE_Pd_fig} (`flares+background' gray line)
together with the observation (`flares+background' black line).
The errors of ($a_0, \nu, \tau / T$) are not independent of each other, and the values
given in equation (\ref{results_d}) are projections of the 
90\% confidence domain (see Fig. \ref{EUVE_chi2_fig} below).

\begin{figure}[h!]
\plotone{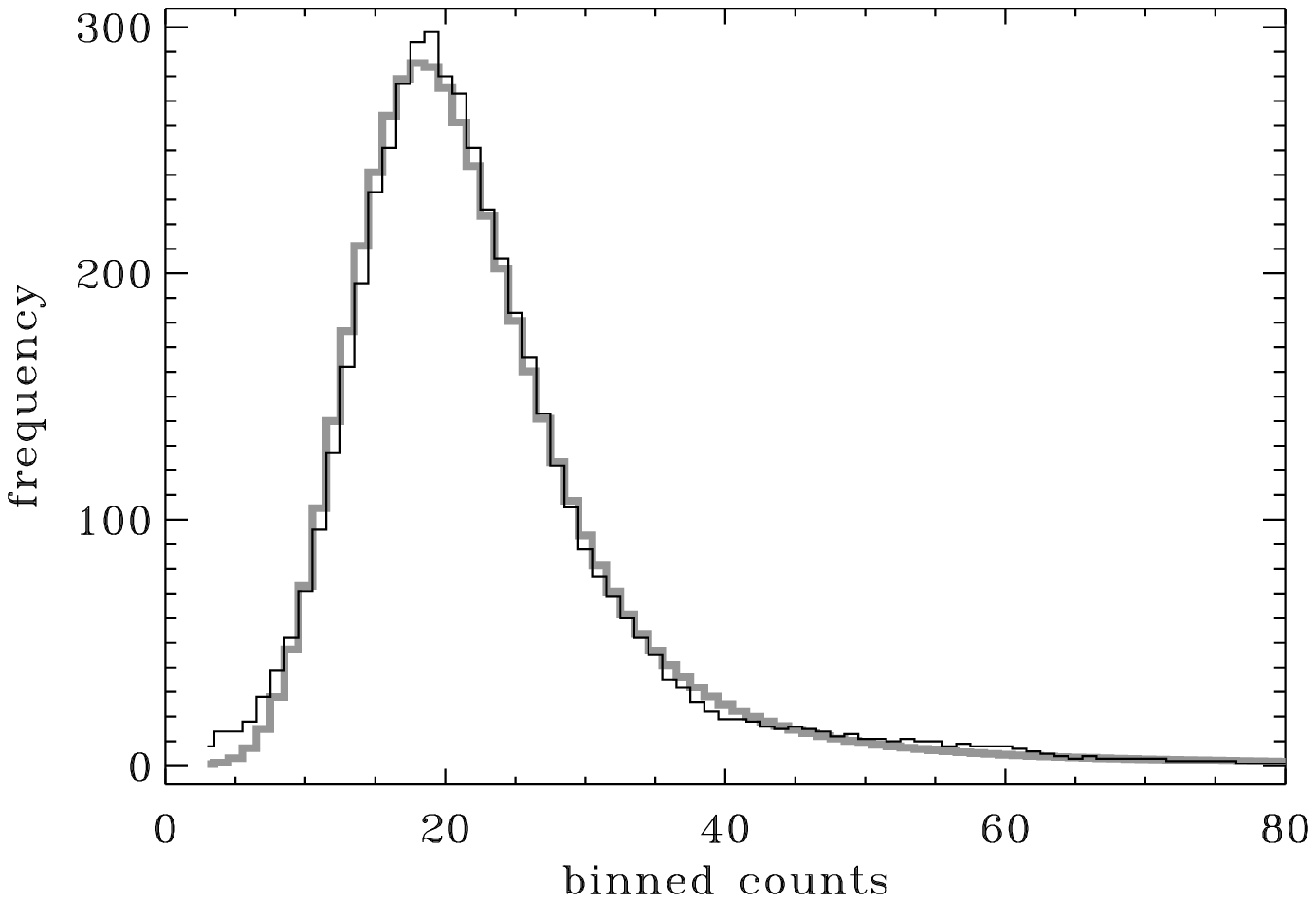}
\caption{Observed (black) and theoretical (gray) histograms of
the binned counts of the EUVE/DS AD Leo observation of Figure \protect\ref{EUVE_lc_fig}.
The time bin size is $\Delta t$ = 100 s.}
\label{EUVE_Pc_fig}
\end{figure}

We also estimated the parameters ($a_0, \nu, \tau / T$) from the observed histogram of binned
counts (Fig. \ref{EUVE_Pc_fig} black line). The (disjoint, adjacent) time bins have duration 
$\Delta t$ = 100 s and contain,
on average, 22 counts. Time bins which would intersect the transition from or to an unobserved time
interval are rejected. In order to abate the arbitrariness of choosing a binning offset,
the starting time of the first time bin has been varied at random within $\Delta t$, and 
the histogram of Figure \ref{EUVE_Pc_fig} is an average over 50 random offsets. Time bins with less
than 2 counts and more than 80 counts are excluded, because they are 
affected by the binning offset and have poor statistics.
The best-fit solution is again found by minimizing the chi 
square expression of equation (\ref{chi2}), where $H_n$ and $P_n$ are
now the observed and theoretical frequencies of $n$ counts in time bins of duration $\Delta t$.
The theoretical prediction is $P_n = N P_c(n)$ with $N$ = $\sum_n H_n$ = 4959 and $n$ running
over $N_o$ = $80 - 2$ = 78 photon number bins.
Since $\Delta t$ $\ll$ $\tau$, we may use the short-term form of the probability 
density of binned counts, $P_c(n) = (2\pi)^{-1} \int_0^{2\pi} \hspace{-2mm} ds \, e^{-ins-bz(s)\Delta t} \phi_f(i z(s) \Delta t)$
with $z(s)=1-e^{is}$ and $\phi_f(s)$ given by equation (\ref{phi_f_pl1}). The minimum reduced chi square (per $N_{\rm DOF}$ 
= $N_o$ $-$ 3 = 75 degrees of freedom) is 1.003, which is attained at
\begin{equation}
\nu = 2.31 \pm 0.12 \;\;\;\;\;\; a_0=0.0028 \pm 0.0015 \; \mbox{ct~$s^{-1}$} \;\;\;\;\;\; \tau / T = 18 \pm 7 \, .
\label{results_c}
\end{equation}
The best-fit binned count distribution is superimposed in Figure \ref{EUVE_Pc_fig} (gray line). 
The errors given in equation (\ref{results_c}) are again projections of the 90\% confidence domain.

\begin{figure}
\plotone{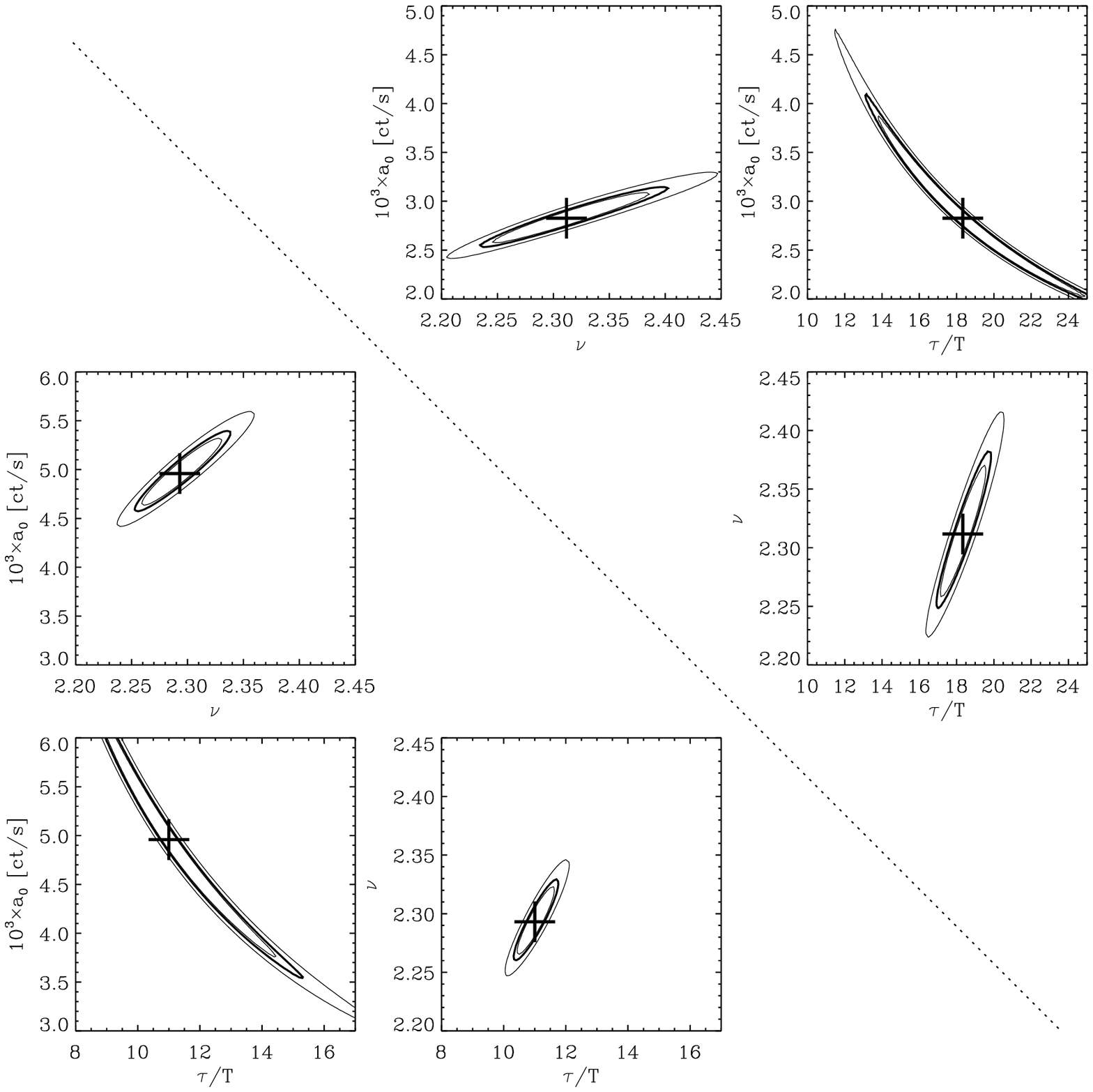}
\caption{Chi square sections in parameter space ($a_0,\nu,\tau / T$) through the
best-fit solution (crosses). Panels above the diagonal refer to the binned count method;
panels below the diagonal refer to the photon waiting time method. The contours represent
(0.75, {\bf 0.9}, 0.999) confidence levels.}
\label{EUVE_chi2_fig}
\end{figure}

The confidence domains of ($a_0, \nu, \tau / T$) for both the photon waiting time and binned 
count methods are visualized in Figure \ref{EUVE_chi2_fig}. The contours represent chi square
sections through the best fit solution (crosses) at levels $\chi^2_{\rm min} + (1.63, 2.33, 4.66)
\sqrt{N_{\rm DOF}}$, corresponding to (0.75, 0.9, 0.999) confidence levels\footnote{If the model
($a_0, \nu, \tau / T$) was true then the probability that the observed $\chi^2$ 
(eq. \protect\ref{chi2}) deviates from its expectation value 
$N_{\rm DOF}$ by more than $\epsilon$ is about $1-{\rm erf} (\epsilon/2\sqrt{N_{\rm  DOF}})$.
The cited `confidence levels' represent the
probabilities that this does not happen.}. Panels above the diagonal refer to the
binned count method; panels below the diagonal refer to the photon waiting time method.
The formal 90\% confidence level is marked boldface. We find that the relative error of $\nu$ is 
smaller than the errors of $a_0$ and $\tau / T$ for both the binned count and waiting time methods. 
The power law index $\nu$ can thus be estimated more reliably than the lower cutoff. The shape
of the confidence domains indicates that the parameters are not independent of each other. 
The strongest interference is between the lower cutoff $a_0$ and the separation of individual flares, 
$\tau/T$, which reflects the fact that the
average flare photon flux, $\tau T^{-1} \langle a \rangle $ = 
$\tau T^{-1} a_0 (1-\nu)/(2-\nu)$, is sensitive to the product $\tau T^{-1} a_0$
only. Likewise, $\nu$ is positively coupled to both $a_0$ and $\tau / T$ because
the function $(\nu-1)/(\nu-2)$ decreases with increasing $\nu$ ($\nu$ $>$ 2).
The results of the photon waiting time (eq. \ref{results_d}) and binned count (eq. \ref{results_c}) 
methods agree within the 90\% errors, which is considered as satisfactory,
although the binned count method systematically
yields lower $a_0$ and larger $\tau / T$ than the photon waiting time method, whereas
the product $\tau T^{-1} a_0$ is similar in both cases (0.0541 for the waiting time method;
0.0527 for the binned count method).

The mean flaring interval $T$ and flare duration $\tau$ enter the short-term forms of
binned count- and photon waiting time distributions only in the dimensionless 
combination $\tau / T$. The decay constant $\tau$ of flares with amplitudes $\ge$ $a_0$
has been investigated by \cite{kashyap02} 
and in \cite{guedel03} and found to be about 3000 s; adopting this value, we find
a mean flaring interval of 
\begin{equation}
T \sim 220 \; \mbox{s}
\label{flare_rate}
\end{equation}
(average over both methods). The use of the short-term forms of the binned count- and photon waiting 
time distributions is justified, a posteriori, since $\Delta t$ $\ll$ $\tau$, and since $\delta t$ $\le$ 30 s $\ll$ $\tau$ 
and $\langle a \rangle \tau$ $\sim$ 20 $\gg$ 1. In fact, direct numerical evaluations of the exact forms
(eqns. \ref{P_c}, \ref{P_d}) at best-fit parameters (eqns. \ref{results_d}, \ref{results_c}, \ref{flare_rate}) deviate 
from their short-term approximations by less than 2\%.

\section{\label{Discussion_Sect}DISCUSSION}

We have applied an analytical approach to the inversion of light curves to
derive the amplitude distribution of constituent elementary flare light curves.
The method uses either binned count light curves or the distribution of
waiting times derived from photon lists. We have in particular considered
flare shapes described by a one-sided exponential decay or a two-sided exponential
profile, and we have studied flare amplitude distributions described by
an exponential function and a truncated power law. The latter
form has been favored in previous studies because solar flare
statistics has revealed power-law distributions over a wide range in energy
\citep{datlowe74, lin84, kruckerbenz98}; explicit stellar studies indicate power laws
as well \citep{collura88, audard99, audard00}. We have thus used a power law to model
EUVE/DS observations of AD Leo, where we find a power law index of $\nu$ = 2.3 $\pm$ 0.1,
a lower cutoff of $a_0$ = 0.004 $\pm$ 0.002 ct~$s^{-1}$, and that there are
$\tau / T$ = 15 $\pm$ 8 flares per flare duration above $a_0$; these values are compilations
of the binned count- and photon waiting time results. 

Our results agree with complementary numerical methods recently applied
to similar data sets \citep{audard99,audard00,kashyap02,guedel03}
indicating power-law indices for the flare amplitude distribution of typically 2.0 -- 2.4,
with most measurements concentrating around $\nu$ = 2.2 -- 2.3. While such values
are necessary for power-law flare amplitude distributions to `explain'
coronal heating on active stars, they are not sufficient. The emitted power
observed from magnetically active coronae can only be due to the ensemble of
flares {\it if} the power-law distribution extends to sufficiently low
energies; this conjecture is of course difficult to prove due to the strong
superposition of near-simultaneous flares that cannot be separated spatially,
thus forming a pseudo-continuous emission level. When expressed in terms of flare energy, 
our numerical results (eqns. \ref{results_d}, \ref{results_c}) correspond to a flare peak luminosity 
and minimum flare energy of
\begin{equation}
a_0 \, \hat{=} \, (0.6 \, ... \, 1.6) \times 10^{27} \; {\rm erg/s} \;\;\;\;\;\;\; 
a_0 \tau \, \hat{=} \, (1.8 \, ... \, 4.8)  \times 10^{29} \; {\rm erg} \, ,
\label{results_erg}
\end{equation}
where it is assumed that one EUVE/DS count corresponds to $2.8 \times 10^{29}$ erg emitted at AD Leo
(see \citealt{guedel03}). Alternatively, a fraction of the observed emission may be truly steady (or variable on time 
scales much longer than flare time scales), but this level should then be below the lowest level seen
in light curves between individual detected flares; \citet*{audard03} estimate
that a mere 30\% (or less) of the low-level emission of the dMe (binary) star
UV Ceti could be ascribed to quasi-steady emission, the rest being recorded as
variable on time scales of tens of minutes with shapes indicating continuous flaring.
Additional information contained in the differential
emission measure distribution could help interpreting the physical processes
participating in the heating of the coronal plasma \citep{guedel03}.

As mentioned in \cite{guedel03}, a few words of caution are in order. A number
of features cannot be recognized in the data per se and would require additional
information. Specifically, the observations collect emission from across
the visible hemisphere of the star, although heating mechanisms or, in our
picture, flare amplitude distributions, may vary across various coronal structures;
further, some of the variability may be introduced by intrinsic changes in
the large-scale magnetic field distributions; emerging or disappearing active
regions will certainly contribute to variability, as do changes in the visibility of
individual regions on the star as they move into or out of sight due to the star's
rotation. Nevertheless, the light curve characteristics do not appear to
be dominated by any of these longer-term effects, while variability on flare time scales is
ubiquitous; similar conclusions were drawn by \citet{audard00}, \citet{kashyap02}, \citet{guedel03},
and \citet{audard03}. We therefore conclude that the present study further supports
the view that stochastic flaring is of principal importance to coronae of magnetically
active stars, and that coronae of such stars may significantly, if not entirely,
be heated by flares.
  
A practical benefit of the analytical method lies in its efficient assessment of the
compatibility of model and observations. The computation of $\chi^2$ for a given
set of parameters takes milliseconds on a medium-size work station, while
a Monte Carlo simulation of comparable accuracy takes seconds or minutes due to slow
($\sim N^{-1/2}$) convergence at rare large flares. The efficiency of the analytical
approach allows to systematically explore the whole parameter space
for the global chi square minimum. For instance, Figure \ref{EUVE_chi2_fig} has been obtained by a brute-force 
calculation of $\chi^2(a_0,\nu,\tau/T)$ on a cubic lattice of dimension 90$\times$95$\times$100, taking about one 
hour computation time, whereas a similar Monte Carlo simulation would take many weeks. 
The possibility for systematic exploration of the parameter space (at low
resolution) is important because the chi square function has generally several local minima, so that
traditional minimization routines such as conjugate gradients may fail.
Note that the simple contours of Figure \ref{EUVE_chi2_fig} depict only the local behavior around 
the global minimum. 

When comparing the binned counts- and photon waiting time methods it was found that the binned counts 
method has larger confidence domains than the photon waiting time method, but is
also more robust in finding the global minimum. A good strategy is thus to start
with the binned counts, using a low-resolution
exploration of the parameter space. As soon as a the approximate global chi
square minimum is found, this may be refined by the photon waiting time
method. In any case the final solutions should be cross-checked for
both methods, to ensure consistency and to obtain an error estimate for the parameters which goes 
beyond the errors of the separate methods.

We would finally stress that our analytical method is in no way restricted to the particular
time series studied in this paper. The method applies whenever the process can be described
in the form of equation (\ref{model}), which may include models for so divers phenomena
as cataclysmic variables, Gamma Ray Burst substructures, and X Ray binaries. There is also
no need to restrict the process (\ref{model}) a the real line (see below). A drawback of 
the analytical method, compared to Monte Carlo simulations, is its limited capability
to incorporate systematic instrumental artifacts. We close by mentioning a few straight 
forward extensions of the model (eq. \ref{model}).

{\bf Flare clustering.} The assumption of independent flare times might be questionable 
in the light of more advanced avalanche models \citep{Vlahos} where flares may trigger each other. Flare clustering can be accounted for
by a flare waiting time distribution
which does not create uniform flare times ($P_\eta(x) \not= T^{-1} e^{-x/T}$), but
has a long-range tail permitting rare but large jumps. The `clusters' are between the jumps.
If the flare amplitudes are still independently $P_a$-distributed, rise immediately and decay 
exponentially, then the functional Equation (\ref{phi_eq}) for the peak values still holds, 
but the separation into the form (\ref{int_eq}) is no longer possible.
However, Equation (\ref{phi_eq}) can be iterated: the sequence
\begin{eqnarray}
\phi_{r_n}^{(0)}(s) 	& = & \phi_a(s) \nonumber \\
\phi_{r_n}^{(k+1)}(s) 	& = & \phi_a(s) \int_0^\infty d x \, P_\eta (x) \, \phi_{r_n}^{(k)}(s e^{-x/\tau})
\label{iter}
\end{eqnarray}
converges uniformly to the characteristic function of $r(t_n+0)$
of clustered flares. All iterates are characteristic functions, and the 
characteristic function of the flare photon intensity is given by
$\phi_f(s) = \lim_{k\to\infty} \phi_{r_n}^{(k)}(s) / \phi_a(s)$ (see Section \ref{Peak_Sect}).

{\bf CCD movies.} The arguments leading to Equations (\ref{phi_f}) and (\ref{phi_F}) are not restricted
to one-dimensional light curves. A useful generalization is to a sequence of spatially
resolved CCD images od solar flares, where binning is over the
space-time pixel $\Delta = \Delta {\bf x} \times \Delta t$, and
the auxiliary interval $I$ of Section 3.1 becomes $I = L^2 \times T$. Provided that the flares occur
independently and uniformly in $I$, the derivation of equation (\ref{phi_F}) 
goes through without modifications if $\Delta t$ is replaced by $\Delta$ and $\Xi(t, \Delta t)$ is
replaced by $\Xi(x,\Delta) = \int_\Delta \xi(x-x') \, dx'$ with $x = ({\bf x},t)$ and $\xi(x)$ the 
space-time flare shape.

\acknowledgments
We thank Marc Audard for discussions on the EUVE data set.

\begin{appendix}

\section{\label{limit_Appendix}The limit $|I| \to \infty$}

By definition, $\phi_a(s)$ satisfies $\phi_a(0)=1$ (normalization).
Let us assume that $\phi_a(s) \sim 1 - c |s|^{\alpha} {\rm sign}(s)$ as $|s| \to 0$ with $\alpha > 0$ and $c>0$;
this is a valid form of $\phi_a(s)$ which is realized, for $\alpha<2$, in L\'evy distributions \citep{Gnedenko-Kolmogorov}. Since $\xi(t) \to 0$ as 
$|t| \to \infty$, the integral in equation (\ref{phi_fT}) remains well-behaved for all $s$ when $|I| \to \infty$ if
$\int_{-\infty}^\infty (\xi(t))^\alpha \, dt < \infty$, requiring that $\xi(t)$ decays faster than $|t|^{-1/\alpha}$ as $|t| \to \infty$. 
So far about the decay of the flare shape. The decay of the probability density $P_a(x)$ at large flare amplitudes $x$ 
is related by Tauberian theorems \citep{stein99} to the small-$s$ limit of the characteristic function. For our assumption
$\phi_a(s) \sim 1 - c |s|^\alpha {\rm sign}(s)$, one has that $P_a(x) = (2\pi)^{-1} \int_{-\infty}^\infty ds \, e^{-isx} \phi_a(s) 
\sim x^{-1-\alpha}$ as $x \to \infty$. These are the conditions mentioned in Section \ref{flare_contrib_Sect}.

\section{\label{explicit_Appendix}Explicit calculations}

For special flare shapes and amplitude distributions, equations (\ref{phi_f}) and (\ref{phi_F})
can explicitly be evaluated. To this end, it is advantageous to cast them in the Volterra form
\begin{eqnarray}
\phi_f(s) & = & \exp \Big\{ - T^{-1} \sum_l \int_0^s d y \, k_l(s,y) (1-\phi_a(y)) \Big\} \label{Volterra_f} \\
\phi_F(s,\Delta t) & = & \exp \Big\{ - T^{-1} \sum_l \int_0^{s \max \Xi (t,\Delta t)} \hspace{-16mm} dy \, 
	K_l(s,y) (1-\phi_a(y)) \Big\} \label{Volterra_F}
\end{eqnarray}
with kernels
\begin{equation}
k_l(s,y) = \frac{1}{s|\xi'(t_l)|} \;\;\;\;
K_l(s,y) = \frac{1}{s |\Xi'(t_l,\Delta t)|} \label{kernel} \, ,
\end{equation}
where $t_l$ is the $l$-th solution of $\xi (t) = y/s$ (kernel $k_l(s,y)$) or $\Xi (t,\Delta t)=y/s$
(kernel $K_l(s,y)$), respectively, and prime denotes derivative with respect to $t$. 
The success of an explicit evaluation of equations
(\ref{Volterra_f},~\ref{Volterra_F}) depends on the simplicity of $\phi_a(s)$  and on the ability to 
invert the functions $\xi (t)$ and $\Xi (t,\Delta t)$ for the kernels $k_l(s,y)$ and $K_l(s,y)$. $\phi_F(s,\Delta t)$
is generally harder to obtain than $\phi_f(s)$.

If the flare shape is $\xi (t) = \theta(t) e^{-t/\tau}$ then the equation $\Xi (t,\Delta t)=y/s$ has the
solutions $t_1$ = $-\tau \ln (1-y/s\tau) - \Delta t$ (valid for the rise phase $-\Delta t < t_1 <0$) 
and $t_2$ = $-\tau \ln (y / s\tau) + \tau \ln (1 - e^{-\Delta t/\tau})$ (valid for the decay phase $t_2 > 0$).
The corresponding kernel contributions (eq. \ref{kernel}) are $K_1=\tau/(s\tau-y)$ and $K_2=\tau/y$,  
and equation (\ref{Volterra_F}) becomes
\begin{equation}
\phi_F(s,\Delta t) = \exp  - \frac{\tau}{T} \int_0^{s\tau(1-e^{-\Delta t/\tau})} \hspace{-15mm} dy \,
	\big( 1-\phi_a(y) \big) \Big(\frac{1}{s\tau-y} + \frac{1}{y} \Big) .
\label{expdecay1}
\end{equation}
The mean of $F(0,\Delta t)$ is $T^{-1} \langle a \rangle \tau \Delta t$, and its variance
is $T^{-1} \langle a^2 \rangle \tau^2 \, (\Delta t -\tau - \tau e^{-\Delta t/\tau})$. 
For short time bins ($\Delta t \ll \tau$) one has that $y \ll s \tau$, and
equation (\ref{expdecay1}) reduces to $\phi_F(s,\Delta t) = \phi_f(s \, \Delta t)$ with
$\phi_f(s)$ given by equation (\ref{phi_f1}). We present now several exactly solvable
models for the flare amplitude distribution.

For $P_a(x) = \theta(x) a^{-1}e^{-x/a}$ and 
one-sided exponential flare shape, all characteristic functions and the photon waiting 
time distribution can be rigorously calculated. The results are
\begin{eqnarray}
\phi_a(s)  & = & (1-ias)^{-1} \\
\phi_f(s)  & = & (1-ias)^{-\tau/T} \label{phi_f_exp} \\
\phi_F(s,x)& = & \Big((1-isa\tau)e^{x/\tau} + isa\tau \Big)^{- \frac{s T^{-1} \tau^2 a}{s \tau a + i}} \label{phi_F_exp}
\end{eqnarray}
%
%
Equation (\ref{phi_f_exp}) indicates that $P_f(x)$ is a chi square distribution with $\tau / T$ (non-integer)
degrees of freedom. The combination of exponential flare amplitudes and exponential flare profiles has the peculiar
property that $\phi_f(s) = \phi_a(s)$ (and thus $P_f(x)=P_a(x)$) if $\tau / T = 1$.
If a constant background $b$ [ct~$s^{-1}$] is present, the exact (eq. \ref{P_d}) and short-term (eqns. \ref{P_d0}, \ref{P_d1}) 
forms of the photon waiting time distributions are
\begin{eqnarray}
P_\delta(x) & = & \frac{e^{-bx} \Big( (1+a\tau)e^{x/\tau}-a\tau \Big)^{-\frac{a \tau^2}{T+a\tau T}-2}}{b+a \tau T^{-1} } \times \nonumber \\
	& & \times \Big\{ \Big[ \Big( b+\frac{a \tau}{T}+ba\tau \Big)e^{x/\tau} \hspace{-1mm} - ba\tau \Big]^2 
	\hspace{-1mm} + \frac{a^2 \tau e^{x/\tau}}{T} \Big\} \label{P_d_exp_bg} \\[2mm]
P_\delta(x) & \simeq & \frac{e^{-bx}(1+ax)^{-\tau / T}}{b+a\tau T^{-1}} 
	\Big\{ \Big( b + \frac{a\tau T^{-1}}{1+ax} \Big)^2 + \frac{a^2 \tau T^{-1} }{(1+ax)^2} \Big\} \, .\label{P_d_exp_bg0}
\end{eqnarray}
Equations (\ref{P_d_exp_bg}) and (\ref{P_d_exp_bg0}) allow to benchmark (and to be benchmarked by) our Monte Carlo 
simulations (Section \ref{Simulations}). They also provide a measure of validity of the short-term approximation
(eqns. \ref{P_d0}, \ref{P_d1}): for example, for $x$ = $0.33 \tau$, $a\tau$ = 25 photons per flare, $\tau / T$ = 1 
(moderate flare overlap) and $b$ = 0.2$a \tau T^{-1}$ (20\% background), the error of the short-term 
approximation is about 0.6\%.

If $P_a(x)$ is a truncated power law (eq. \ref{P_a_pl})
and if the flare shape is exponential, then $\phi_f(s)$ may be expressed in 
terms of special functions. Possible representations are
\begin{eqnarray}
\ln \phi_f(s) & = & \frac{\tau}{T} \Big\{ \frac{\pi (-ia_0s)^{\nu-1}}{\Gamma(\nu)\sin(\nu \pi)} + ia_0s \frac{\nu-1}{\nu-2} 
		\; {_3F_3}([1,1,2-\nu],[2,2,3-\nu],ia_0s)  \Big\} \label{phi_f_pl1} \\
\phi_f(s) & = & e^{\frac{\tau}{T}(\gamma - \frac{i\pi}{2} + \frac{1}{\nu-1})} \, (a_0s)^{-\tau/T} 
		e^{-\frac{\tau}{T}\big\{ {\rm E_1}(-ia_0s) + \nu (-ia_0s)^{\nu-1} \Gamma(-\nu,-ia_0s) 
		+ \frac{e^{ia_0s}}{ia_0s} \big\} } \label{phi_f_pl2}
\end{eqnarray}
%
%
where $_pF_q({\bf n}, {\bf d}, z)$ denotes the generalized hypergeometric function \citep{Gradshteyn},
$\gamma=0.5772$ is Euler's constant, $\Gamma(a,z)$ is the incomplete gamma function \citep{Abramowitz},
${\rm E_1}(z) = \Gamma(0,z)$ is the exponential integral, and the principal branches of the $_qF_q$ and
$\Gamma$ functions are to be used. Equations (\ref{phi_f_pl1}, \ref{phi_f_pl2}) 
have been obtained by the computer algebra program MAPLE.
In numerical computations we use the form (\ref{phi_f_pl1}) 
with a Taylor (small-argument) expansion of the hypergeometric function. 
For values of $a_0s$ $\ll$ 1, this expansion converges rapidly, and machine precision is 
reached after $<$ 6 orders if $a_0s$ $\le$ 0.25. In fact, already order two yields $<$ 0.0005 rms relative precision for $P_\delta(x)$
over the whole waiting time range 0 $\le$ $\delta t$ $\le$ 30s used in the EUVE data. 
This fact is exploited in the approximation (\ref{Pd_approx}), which is based on an order-two
Taylor expansion of the hypergeometric function.

If $P_a(x)$ is a Le\'vy distribution \citep{Levy, Gnedenko-Kolmogorov} with characteristic
function $\phi_a(s) = \exp \{ -c|s|^\alpha ( 1-i\frac{s}{|s|} \frac{\pi \alpha}{2}) \}$, $c>0$,
and 0 $<$ $\alpha$ $<$ 1, then $P_a(x)$ vanishes at negative $x$ and therefore can
model flare amplitudes. Large flare amplitudes ($x$ $\gg$ $c$) decay like $P_a(x)$ $\sim$ $x^{-1-\alpha}$, and
superimposed flares have $\phi_f(s)$ = $(e^{\gamma} \eta)^{-\lambda \tau / \alpha} |s|^{-\lambda \tau} 
\exp \{ -\frac{\lambda \tau}{\alpha} {\rm {\rm E_1}} (\eta |s|^\alpha)\}$
with $\eta = c \big( 1-i \frac{s}{|s|} \tan \frac{\pi \alpha}{2} \big)$ and
$\gamma$ is Euler's constant (see above). Such L\'evy distributions might be a convenient model
for small solar flares where power law indices $<$ 2 have been reported \citep{crosby93}.

\section{\label{moment_Appendix}Alternative Approach: Moments and Cumulants}

We ask for an elementary derivation of the results (\ref{phi_f} - \ref{P_d}).
The key observation is that equations (\ref{phi_f}) and (\ref{phi_F}) imply a simple
relation between the moments of the flare amplitudes and 
the cumulants\footnote{The cumulants of a quantity $x$ are defined by $\kappa_n(x) 
= (-i)^n \partial_s^n \ln \phi_x(s) |_{s=0}$. Cumulants of order $\le n$ can be 
expressed in terms of moments of order $\le n$, and vice versa; see, e.g.,
\cite{Eadie71}.} of the instantaneous photon intensity and the binned Poisson 
parameter.  This fact may be exploited to establish a
relation between the moments of the flare amplitudes and the moments of the 
binned counts and photon waiting times, without recurrence to
characteristic functions.
We illustrate the procedure starting with the expectation value $\langle f(0) \rangle$.
Interpreting the angular brackets as time average (which is justified by the
assumption of stationarity), the central limit theorem ensures that
\begin{equation}
\langle f(0) \rangle = \lim_{|I|\to \infty} \frac{1}{|I|} \int_I dt \sum_{t_k \in I} a_k \xi(t-t_k)
= \langle a \rangle \lim_{|I|\to \infty} \frac{\langle {\cal N}_I\rangle}{|I|} \hspace{-1mm}
	\int_I dt \, \xi(t) + O \Big({\cal N}_I^{1/2}\Big) \to \langle a \rangle \frac{1}{T} 
	\hspace{-1mm} \int \hspace{-1mm} \xi(t) \, dt
\label{time_mean}
\end{equation}
where ${\cal N}_I$ is the number of flares in the interval $I$ (see Section \ref{flare_contrib_Sect}). 
When restricting the first sum in eq. (\ref{time_mean}) to flares inside $I$ it has been 
supposed that contributions from 
outside $I$ decay sufficiently fast to be neglected, and we have used the fact that all $a_k$ and $t_k$
are statistically independent of each other. Similarly, the second moment of $f(0)$ is given by
\begin{eqnarray}
\langle f(0)^2 \rangle & = & \lim_{|I|\to \infty} \frac{1}{|I|} \int_I dt 
		\Big(\sum_{t_k \in I} a_k \xi(t-t_k)\Big) \, \Big(\sum_{t_l \in I} a_l \xi(t-t_l)\Big) \nonumber \\
	& = & \bigg\{ \sum_{k\not=l} + \sum_{k=l} \bigg\} \langle a_k a_l \rangle 
		\lim_{|I|\to \infty} \frac{1}{|I|} \int_I dt \, \langle  \xi(t-t_k) \xi(t-t_l) \rangle  \label{kl_split} \\
	& = & \bigg\{ \sum_{k\not=l} + \sum_{k=l} \bigg\} \langle a_k a_l \rangle 
		\lim_{|I|\to \infty} \frac{1}{|I|} \int_I dt 
		\int_I \frac{dt_k}{|I|} \int_I \frac{dt_l}{|I|} \xi(t-t_k) \xi(t-t_l)  \nonumber \\
	& = & \langle a \rangle^2 \frac{\langle {\cal N}_I^2 - {\cal N}_I \rangle}{|I|^2} 
		\Big( \int \hspace{-1mm} \xi(t) \, dt \Big)^2
		+ \langle a^2 \rangle \frac{\langle {\cal N}_I \rangle}{|I|} \int [\xi(t)]^2 \, dt 
		+ O\Big({\cal N}_I^{1/2}\Big)  \nonumber \\
	& \to & \langle a \rangle^2 T^{-2} \Big( \int \hspace{-1mm} \xi(t) \, dt \Big)^2 
		+ \langle a^2 \rangle T^{-1} \int [\xi(t)]^2 \, dt \label{time_var} \, .
\end{eqnarray}
From equations (\ref{time_mean}) and (\ref{time_var}) we recognize 
that $\langle f(0)^2 \rangle 
- \langle f(0) \rangle^2 = \langle a^2 \rangle T^{-1} \int [\xi(t)]^2 \, dt$: the
variance of the instantaneous photon intensity equals the
second moment of the flare amplitudes times a weight $\int [\xi(t)]^2 \, dt$.
This finding can be generalized to higher cumulants if the division
into $k$=$l$ and $k$$\not=$$l$ contributions in equation (\ref{kl_split}) is 
replaced by a sum over partitions into equal and non-equal indices. We shall
not discuss here the (complicated) combinatorial weights, but note that they 
correspond to those of the cumulants expressed in
terms of moments (e.g., \citealt{rota00}). As a consequence,
\begin{equation}
\kappa_n(f) = \langle a^n \rangle T^{-1} \hspace{-1mm} \int [\xi(t)]^n \, dt \, .
\label{cumulants_f}
\end{equation}
Equation (\ref{cumulants_f}) is a weak form of equation (\ref{phi_f}), since it
represents its Taylor expansion around $s$=0. It is easily seen that equation
(\ref{cumulants_f}) also applies to the binned Poisson parameter $F(0,\Delta t)$ if 
$\xi(t)$ is replaced by $\Xi(t,\Delta t)$. Therefore the cumulants (and thus the moments) of 
both $f(0)$ and $F(0,\Delta t)$ can be expressed in terms of the moments of the flare 
amplitudes. The probability density of the binned counts $n$ is then given by
$P_c(n) = \langle F(0,\Delta t)^n e^{-F(0,\Delta t)} \rangle / n! = (n!)^{-1} \sum_{k=0}^\infty (-1)^k 
\langle F(0,\Delta t)^{n+k} \rangle /k!$, while the probability density of short 
(Section \ref{WaitingTimeSect}) photon waiting times $x$ is $P_\delta(x) \propto 
\langle f(0)^2 e^{-x f(0)} \rangle = \sum_{k=0}^\infty (-x)^k \langle f(0)^{2+k} \rangle / k!$.
Although these series may be evaluated in special cases, it is generally much more
convenient to work with the characteristic functions, which collect all the moments  
in a compact way.

\end{appendix}

\end{document}